\documentclass[12pt]{iopart}
\expandafter\let\csname equation*\endcsname\relax
\expandafter\let\csname endequation*\endcsname\relax
\usepackage{dcolumn}
\usepackage{bm}
\usepackage{graphics}
\usepackage{longtable}
\usepackage{units}
\usepackage{textcomp}
\usepackage{float}
\usepackage{mathtools}
\usepackage{amsfonts}
\usepackage[font=footnotesize]{caption}
\usepackage{titlesec}

\titleformat{\section}
{\normalfont\large\bfseries}{\thesection}{1em}{}
\titleformat{\subsection}
{\normalfont\normalsize\bfseries}{\thesubsection}{1em}{}
\titleformat{\subsubsection}
{\normalfont\small\bfseries}{\thesubsubsection}{1em}{}

\makeatletter
\newsavebox{\@brx}
\newcommand{\llangle}[1][]{\savebox{\@brx}{\(\m@th{#1\big\langle}\)}%
	\mathopen{\copy\@brx\mkern2mu\kern-0.9\wd\@brx\usebox{\@brx}}}
\newcommand{\rrangle}[1][]{\savebox{\@brx}{\(\m@th{#1\big\rangle}\)}%
	\mathclose{\copy\@brx\mkern2mu\kern-0.9\wd\@brx\usebox{\@brx}}}
\makeatother

\DeclareMathOperator{\sinc}{sinc}

\begin{document}

\title{Second-order correlations in single-particle interferometry}

\author{A Rembold$^1$, R R\"{o}pke$^1$, G Sch\"{u}tz$^1$, J Fort\'{a}gh$^2$, A Stibor$^{1,+}$ and A G\"{u}nther$^{2,*}$}
\address{$^1$ Institute of Physics and Center for Collective Quantum Phenomena in LISA$^+$,
University of T\"{u}bingen, Auf der Morgenstelle 15, 72076 T\"{u}bingen, Germany}
\address{$^2$ Institute of Physics and Center for Collective Quantum Phenomena in LISA$^+$,
University of T\"{u}bingen, Auf der Morgenstelle 14, 72076 T\"{u}bingen, Germany}
\ead{$^+$alexander.stibor@uni-tuebingen.de and $^*$a.guenther@uni-tuebingen.de}

\begin{abstract}
Interferometers with single particles are susceptible for dephasing perturbations from the environment, such as electromagnetic oscillations or mechanical vibrations. On the one hand, this limits sensitive quantum phase measurements as it reduces the interference contrast in the signal. On the other hand, it enables single-particle interferometers to be used as sensitive sensors for electromagnetic and mechanical perturbations. Recently, it was demonstrated experimentally, that a second-order correlation analysis of the spatial and temporal detection signal can decrease the electromagnetic shielding and vibrational damping requirements significantly. Thereby, the relevant matter-wave characteristics and the perturbation parameters could be extracted from the correlation analysis of a spatially ``washed-out" interference pattern and the original undisturbed interferogram could be reconstructed. This method can be applied to all interferometers that produce a spatial fringe pattern on a detector with high spatial and temporal single-particle resolution. In this article, we present and discuss in detail the used two-dimensional second-order correlation theory for multifrequency perturbations. The derivations of an explicit and approximate solution of the correlation function and corresponding amplitude spectra are provided. It is explained, how the numerical correlation function is extracted from the measurement data. Thereby, the influence of the temporal and spatial discretization step size on the extracted parameters, as contrast and perturbation amplitude, is analyzed. The influence of noise on the correlation function and corresponding amplitude spectrum is calculated and numerically cross-checked by a comparison of our theory with numerical single-particle simulations of a perturbed interference pattern. Thereby, an optimum spatial discretization step size is determined to achieve a maximum signal-to-noise ratio. Our method can also be applied for the analysis of broad-band frequency noise, dephasing the interference pattern. Using Gaussian distributed noise in the simulations, we demonstrate that the relevant matter-wave parameters and the applied perturbation spectrum can be revealed by our correlation analysis.
\end{abstract}

\section{Introduction}\label{sec1}
Single-particle interferometry became a remarkable tool in various fields of quantum physics and sensor technology. Interferometers for coherent atoms \cite{Cronin2009,Carnal1991a,Keith1991a} recently investigated the nature of time \cite{Margalit2015,Arndt2015a} and measured inertial forces \cite{gustavson1997} and gravitational acceleration \cite{Peters1999}. Molecule interferometers  \cite{Cronin2009,Grisenti2000a} proved the wave nature of large particles \cite{Brezger2002a,Gerlich2011,Arndt2014a,Haslinger2013} and contributed to the understanding of quantum decoherence \cite{Zurek2003,Hackermuller2004,Hornberger2003}. In neutron interferometers \cite{Rauch1974}, the quantum-mechanical phase shift due to the earth's gravitational field was observed \cite{Colella1975}. Moreover, remarkable progress was achieved in the field of matter-wave interferometry with charged particles such as electrons and ions \cite{Hasselbach2010,Hasselbach1998a,Maier1997} based on new developments concerning the beam source \cite{Kuo2006,Ehberger2015,Hommelhoff2006}, the precise electron guiding \cite{Hasselbach1988,Hammer2014}, the coherent beam path separation by nanostructures \cite{Ehberger2015,Schuetz2014,Chang2009,Cho2004} and highly resolved spatial and temporal single-particle detection \cite{Jagutzki2002}. This advance opened the door for experiments in Aharonov-Bohm physics \cite{Aharonov1959,Batelaan2009,Schuetz2015b} and Coulomb-induced decoherence \cite{Zurek2003,Sonnentag2007,Scheel2012}. Technical devices with interfering single particles can decrease the amount of destructive particle deposition in electron microscopy for the analysis of fragile biological specimen \cite{Putnam2009,Kruit2016}.

All these technical \cite{Grattan2013} and fundamental applications \cite{Abbott2016,Graham2013} of single-particle interferometers are based on the high phase sensitivity and are therefore susceptible to dephasing, which can be caused by external electromagnetic oscillations, mechanical vibrations or temperature drifts. In contrast to decoherence \cite{Sonnentag2007,Scheel2012}, where actual information of the quantum state is lost to the environment, dephasing is a collective, time-dependent phase shift of the interference pattern. Both, decoherence and dephasing, cause a reduction of the contrast of the time-averaged interference pattern on the detector. However, in opposite to decoherence, dephasing can in principle be corrected after the measurement, if the temporal and spatial information of the single-particle events are known. Then, two-particle correlations may be used to study the dephasing process and to reveal the undisturbed interference pattern.

Ever since the famous Hanbury Brown and Twiss experiment \cite{brown1956}, second-order correlations are successfully used in many research areas. Thereby, noise correlations play a key role, as they give direct access to the quantum nature of the source. This understanding has not only set the fundament for modern quantum optics \cite{glauber1963}, but also helped to prove the quantum nature of fermions \cite{kiesel2002} and bosons \cite{schellekens2005}. Today, noise correlation analysis is widely used in modern astrophysics \cite{foellmi2009}, quantum atom optics \cite{simon2011} and particle physics \cite{agakishiev2012}. For matter-waves, temporal correlations have been used to analyze the counting statistics of atom lasers \cite{ottl2005} and to demonstrate the coherent transfer of magnetic field fluctuations onto an atom laser \cite{federsel2017}. Spatial correlations, on the other hand, have been used to analyze atoms in optical lattices \cite{grondalski1999} or to study many-body states, such as the Mott insulator state, in cold atom physics \cite{altman2004,folling2005}.

In previous publications \cite{Rembold2014,Guenther2015,Rembold2016} we have demonstrated in a biprism electron interferometer \cite{Hasselbach2010,Schuetz2014,Mollenstedt1956a}, how multifrequency dephasing caused by electromagnetic \cite{Rembold2014,Guenther2015} and vibrational oscillations \cite{Rembold2016} can be corrected using second-order correlation analysis in combination with the amplitude spectrum of the correlation function. Latter can be used for the identification of unknown perturbation frequencies \cite{Rembold2016}, as according to the Wiener-Khintchine theorem \cite{Wiener1930,Khintchine1934} the Fourier transform of the correlation function is equal to the power spectrum of the perturbed measurement signal. For the measurements, an interference pattern was shifted artificially by external perturbations leading to a contrast reduction of the temporally integrated pattern on the detector. Using the time and position information of particle impacts at the detector, the numerical second-order correlation function was extracted. With this, we were able to reveal the unknown perturbation frequencies, corresponding amplitudes and the characteristics of the matter-wave, such as contrast and pattern periodicity. The undisturbed interference pattern could be reconstructed with the parameters of the perturbation. Our method is a powerful tool to prove the wave nature of particles, even if the integrated interference pattern is vanished. Therefore, it reduces the requirements for electromagnetic shielding and vibrational damping of the experimental setup, e.g. for mobile interferometers or experiments in a perturbing environment. Furthermore, it can be used to sensitively detect electromagnetic and mechanical perturbations or for the spectroscopy of the electromagnetic and vibrational response spectrum of an interferometer \cite{Rembold2016}. Therefore, this technique has the potential for the application in sensor technology and can in principle be applied in interferometers for electrons \cite{Jagutzki2002}, ions \cite{Jagutzki2002}, neutrons \cite{Siegmund2007}, atoms \cite{Schellekens2005} and molecules \cite{Zhou2012} that generate a spatial fringe pattern in the detection plane. For the application of the correlation analysis, the devices have to be equipped with a detector with high spatial and temporal single-particle resolution, which is available for all above mentioned interferometers. Another requirement is, that the particle flight time is shorter than the cycle time of the perturbation. Otherwise, the particles traverse many periods of the perturbation and therefore the perturbation is averaged out and can not be resolved.

This article provides a comprehensive description of the applied theory, being the base for the experimental application of second-order correlations in single-particle interferometry. In the first chapter, we give a detailed derivation of the two-dimensional second-order correlation theory for multifrequency perturbations leading to the equations applied in former dephasing experiments \cite{Rembold2014,Guenther2015,Rembold2016}. We deduce the explicit solution for the correlation function and explain under which conditions an approximation can be applied. The characteristics of the explicit and approximate solutions are discussed and the determination of the matter-wave properties is shown. Furthermore, we calculate the analytic solution of the corresponding amplitude spectrum used for the identification of the unknown perturbation frequencies and amplitudes \cite{Rembold2016}. The invariance of the correlation function under time and space transformations is analyzed in detail and the consequence for the determination of the perturbation parameters is shown. 

In the second part of this article, we investigate the general characteristics of numerical correlation functions. They are typically derived from a finite set of measurement data, causing statistics and noise to play an important role. Temporal and spatial discretization, then influences not only the contrast and the amplitude spectrum of the correlation function, but also the corresponding noise levels. This limits the maximal signal-to-noise ratio in the correlation analysis. From our theoretical study, we identify an optimal discretization step size for best sensitivity. Using single-particle simulations of a perturbed interference pattern, we cross-check our theoretical description and show, how the correlation theory can be used to identify broad-band frequency noise.

\section{Theory of second-order correlations}\label{sec2}
In this chapter, the theory of second-order correlations in single-particle interferometry is derived and the properties are discussed in detail. First, the contrast reduction of a time-averaged interference pattern dephased by a perturbation is analyzed. Afterwards, the explicit solution of the second-order correlation function is calculated and discussed under which conditions an approximate solution is suitable. The determination of the contrast and spatial periodicity of the unperturbed interference pattern is demonstrated. The corresponding amplitude spectra used for the identification of unknown perturbation frequencies are derived. At the end of the chapter, the invariance of the correlation function under time and space transformations is analyzed in detail and the consequence for the determination of the perturbation phases is discussed. 

\subsection{Time-averaged interference pattern}\label{sec2.1}
In many experiments in single-particle interferometry, the interference pattern is detected using multichannel plates (MCPs) in conjunction with a phosphor screen \cite{Hasselbach2010,Hasselbach1998a,Hasselbach1988,Sonnentag2007,Hasselbach1993}. The particle impacts generate light pulses on the phosphor screen, that are temporally integrated with a charge-coupled device camera (CCD-camera). An interference pattern, that is dephased by a temporal perturbation, is then irreversibly ``washed-out" in the spatial signal and its contrast is reduced. This behaviour shall be calculated in the following.

\begin{figure}
\centering
\includegraphics[width=0.5\textwidth]{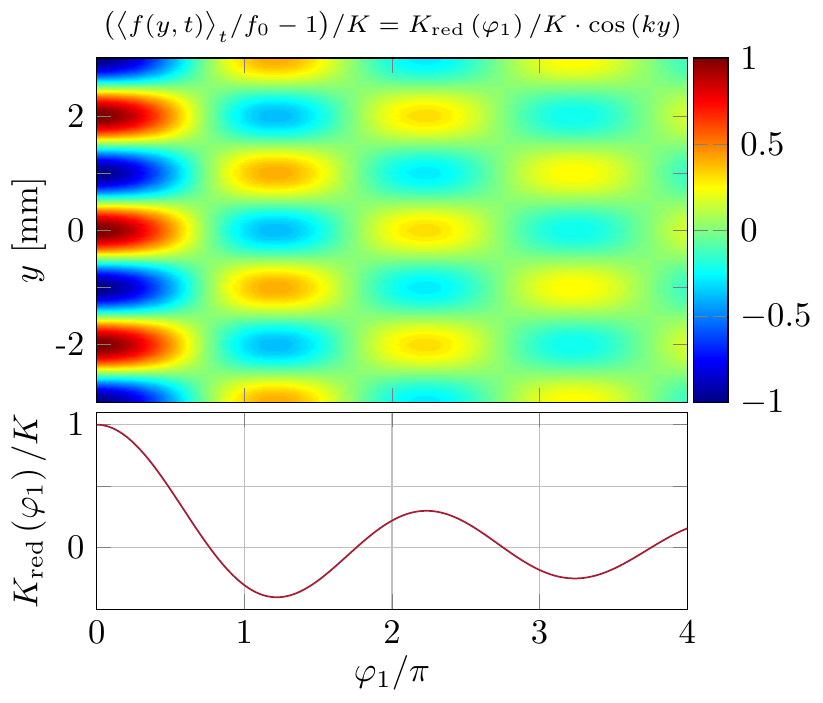} \caption{Top: Dependence of the time-averaged interference pattern (equation (\ref{eq7})) on the peak phase deviation $\varphi_1$ for an interference pattern with a periodicity of $\lambda = \unit[2]{mm}$ along the $y$-direction. Bottom: Dependence of the relative contrast reduction $K_{\text{red}}\left(\varphi_1\right)/K$ on the peak phase deviation $\varphi_1$. For $\varphi_1=\unit[0.76]{\pi}$, the contrast vanishes due to the first zero of $J_0(\varphi_1)$. The contrast returns for larger peak phase deviations, but not completely. At each sign change of $J_0(\varphi_1)$, the phase of the interference pattern is shifted by $\pi$.}
\label{fig1}
\end{figure}

The probability distribution, that describes the particle impacts in the detection plane forming the interference pattern, is given by
\begin{equation}
f(y,t) = f_0\Big(1+K\cos \big(ky + \varphi\left(t\right)\big)\Big) ~,
\label{eq1}
\end{equation}
where $f_0$ assures normalization, $K$ and $k=2\pi/\lambda$ indicate the contrast and wave number of the unperturbed interference pattern, with the spatial periodicity $\lambda$. The time-dependent perturbation $\varphi(t)$ is described as a superposition of $N$ harmonic frequencies $\omega_j$
\begin{equation}
\varphi(t)=\sum_{j=1}^N \varphi_j\cos\left(\omega_j t + \phi_j\right) ~,
\label{eq2}
\end{equation}
with perturbation amplitudes (peak phase deviations) $\varphi_j$ and phases $\phi_j$. The perturbation leads to a washout of the time-averaged interference pattern. For one perturbation frequency ($N=1$), this can be easily seen by calculating the time-average of equation (\ref{eq1})
\begin{equation}
\big\langle f(y,t) \big\rangle_{t} \,\coloneqq\lim_{T\rightarrow\infty} \frac{1}{T}\int_0^T f(y,t)\, \mathrm{d}t ~,
\label{eq3}
\end{equation}
resulting in
\begin{align}\label{eq4}
\big\langle f(y,t) \big\rangle_{t} \,&=\lim_{T\rightarrow\infty} \frac{f_0}{T}\int_0^T\left(1 +\frac{K}{2}\Big(\mbox{e}^{i\left(ky+\varphi(t)\right)}+c.c.\Big)\right)\mathrm{d}t \nonumber\\ 
&= f_0+\lim_{T\rightarrow\infty}\frac{f_0 K}{2T}\left(\mbox{e}^{iky}\int_0^T \mbox{e}^{i\varphi_1\cos(\omega_1 t+\phi_1)}\,\mathrm{d}t+c.c.\right)~.
\end{align} 
Using Bessel functions of first kind $J_n$, the exponential function can be rewritten as
\begin{equation}\label{eq5}
\mbox{e}^{\pm i\varphi_1\cos(\omega_1 t+\phi_1)} = \sum_{n_1=-\infty}^{+\infty}(\pm i)^{n_1}\, \!J_{n_1}(\varphi_1)\mbox{e}^{in_1\left(\omega_1 t+\phi_1\right)}=\sum_{n_1=-\infty}^{+\infty}\, \!J_{n_1}(\varphi_1)\mbox{e}^{in_1\left(\omega_1 t+\phi_1\pm\frac{\pi}{2}\right)}~,
\end{equation}
yielding
\begin{align}\label{eq6}
\big\langle &f(y,t) \big\rangle_{t} \, = \\ \nonumber
&=f_0+\frac{f_0K}{2}\sum_{n_1=-\infty}^{+\infty}\!J_{n_1}(\varphi_1)\bigg(\mbox{e}^{iky}\mbox{e}^{in_1\left(\phi_1+\frac{\pi}{2}\right)}+\mbox{e}^{-iky}\mbox{e}^{in_1\left(\phi_1-\frac{\pi}{2}\right)}\bigg)\lim_{T\rightarrow\infty}\frac{1}{T}\,\int_0^T \mbox{e}^{in_1\omega_1 t}\,\mathrm{d}t~. 
\end{align}
Only for $n_1=0$, the limit of the time integral is equal to one. For all other $n_1\neq 0$, it approaches zero, such that the time-averaged interference pattern becomes
\begin{align}\label{eq7}
\big\langle f(y,t) \big\rangle_{t} \, = f_0\big(1+\underbrace{K\,\!J_0(\varphi_1)}_{=\, K_{\text{red}}\left(\varphi_1\right)}\cos(ky)\big)~.
\end{align}
The perturbation thus leads to a reduced contrast $K_{\text{red}}\left(\varphi_1\right)$ given by the perturbation amplitude: $K_{\text{red}}\left(\varphi_1\right)=KJ_0(\varphi_1)$, with $\left| J_0(\varphi_1)\right| \leq 1 $. In figure \ref{fig1} at the top, the dependence of the time-averaged interference pattern (equation (\ref{eq7})) on the peak phase deviation $\varphi_1$ is illustrated. The interference pattern with a spatial periodicity of $\lambda = \unit[2]{mm}$ can be identified in the $y$-direction. The reduced contrast $K_{\text{red}}\left(\varphi_1\right)$ normalized to the contrast of the unperturbed interference pattern $K$ is plotted at the bottom of figure \ref{fig1}. For a peak phase deviation of $\varphi_1=\unit[0.76]{\pi}$, the contrast is zero corresponding to the first zero of $J_0(\varphi_1)$. The contrast returns for larger peak phase deviations, but does not recover completely. Additionally, the interference pattern is phase shifted by $\pi$ as the sign of $J_0(\varphi_1)$ changes from positive to negative. This behaviour is repeated for higher peak phase deviations and the contrast is further reduced. 

For multifrequency perturbations ($N>1$) with peak phase deviations $\varphi_j<1$, equation (\ref{eq7}) becomes
\begin{align}\label{eq8}
\big\langle f(y,t) \big\rangle_{t} \, = f_0\big(1+\underbrace{K\,\prod_{j=1}^N J_0(\varphi_j)}_{=\, K_{\text{red}}\left(\varphi_{j=1\ldots N}\right)}\cos(ky)\big)~.
\end{align}
Here, the reduced contrast $K_{\text{red}}\left(\varphi_{j=1\ldots N}\right)$ depends on all peak phase deviations $\varphi_j$, via the product of the zeroth order Bessel functions. Therefore, the contrast is typically stronger reduced as compared to the single frequency case in figure \ref{fig1}.

\subsection{Solution for the analytic second-order correlation function}\label{sec2.2}
With the spatial and temporal information of the particles arriving in the detection plane, it is possible to reveal the contrast $K$ and spatial periodicity $\lambda$ of the unperturbed interference pattern by correlation analysis. This is possible, because in the correlation function the spatial and temporal differences are taken into account in contrast to the temporally integrated interference pattern. Then, on timescales below the perturbation frequency, the interference pattern is still visible and not influenced by the perturbation. Furthermore, the characteristics of the perturbation (frequencies $\omega_j$, peak phase deviations $\varphi_j$ and phases $\phi_j$) can be determined from the correlation analysis. 

With the probability distribution in equation (\ref{eq1}), the second-order correlation function reads
\begin{equation}
g^{(2)}(u,\tau) = \frac{\llangle f(y+u,t+\tau) f(y,t)\rrangle_{y,t}}{\llangle f(y+u,t+\tau) \rrangle_{y,t} \llangle f(y,t) \rrangle_{y,t}}~,
\label{eq9}
\end{equation}
with $\llangle \cdot \rrangle_{y,t}$ denoting the average over position and time
\begin{equation}
\llangle f(y,t) \rrangle_{y,t} \,\coloneqq \lim_{Y,T\rightarrow\infty} \frac{1}{TY}\int_0^T \int_{-Y/2}^{Y/2} f(y,t) \,\mathrm{d}y\mathrm{d}t ~.
\label{eq10}
\end{equation}
If the acquisition time $T$ and length $Y$ are large compared to the involved perturbation frequencies $T\gg 2\pi/\omega_j$ and spatial periodicity $Y\gg \lambda$,  equation (\ref{eq9}) can be solved analytically. First, the term $\llangle f(y+u,t+\tau) \rrangle_{y,t}$ in the denominator of equation (\ref{eq9}) is calculated, using equation (\ref{eq1}) and (\ref{eq10})
\begin{align}\label{eq11}
\llangle f(y+u,t+\tau)& \rrangle_{y,t} \,=\nonumber\\
&=\lim_{Y,T\rightarrow\infty} \frac{f_0}{TY}\int_0^T \int_{-Y/2}^{Y/2} \Big(1+K\cos \big(k\left(y+u\right) + \varphi\left(t+\tau\right)\big)\Big) \,\mathrm{d}y\mathrm{d}t \nonumber\\ 
&=f_0+\lim_{Y,T\rightarrow\infty}\frac{f_0K}{TY}\frac{1}{2}\int_0^T \int_{-Y/2}^{Y/2} \Big(\mbox{e}^{i\left(k(y+u)+\varphi(t+\tau)\right)}+c.c.\Big) \,\mathrm{d}y\mathrm{d}t \nonumber\\ 
&=f_0+\lim_{Y,T\rightarrow\infty}\frac{f_0K}{TY}\Bigg(\mbox{e}^{iku} \underbrace{\frac{1}{2}\int_{-Y/2}^{Y/2} \mbox{e}^{iky}\mathrm{d}y}_{=\, \sin\left(k\frac{Y}{2}\right)/k}\int_0^T\mbox{e}^{i\varphi(t+\tau)} \mathrm{d}t+c.c.\Bigg)~.
\end{align}
In the limit of large acquisition length $Y\gg \lambda$, the spatial limit becomes
\begin{equation}\label{eq12}
\lim_{Y\rightarrow\infty}\frac{\sin\left(k\frac{Y}{2}\right)}{kY}\rightarrow 0~,
\end{equation}
yielding
\begin{equation}\label{eq13}
\llangle f(y+u,t+\tau) \rrangle_{y,t} \,=\, f_0~.
\end{equation}
Using $u=0$ and $\tau=0$ in equation (\ref{eq11}), the second term in the denominator becomes $\llangle f(y,t) \rrangle_{y,t} \,=\, f_0$. This is expected, because shifts in time and space should not alter the long time and position average of the probability distribution.

Using equation (\ref{eq1}) and (\ref{eq10}), the numerator in equation (\ref{eq9}) results in
\begin{align}\label{eq14}
\llangle &f(y+u,t+\tau) f(y,t)\rrangle_{y,t}\,=\\ \nonumber
&=\lim_{Y,T\rightarrow\infty} \frac{f_0^2}{TY}\int_0^T \int_{-Y/2}^{Y/2} \Big[1+K\cos \big(k\left(y+u\right) + \varphi\left(t+\tau\right)\big)+K\cos \big(ky + \varphi\left(t\right)\big)+\\ \nonumber
&\qquad\qquad\qquad\qquad\qquad+K^2\cos \big(k\left(y+u\right) + \varphi\left(t+\tau\right)\big)\cos \big(ky + \varphi\left(t\right)\big)\Big] \,\mathrm{d}y\mathrm{d}t~.
\end{align}
Similar as before, the second and third term in equation (\ref{eq14}) vanish, leaving
\begin{align}\label{eq15}
\llangle &f(y+u,t+\tau) f(y,t)\rrangle_{y,t}\,= \nonumber\\ 
&=f_0^2+\lim_{Y,T\rightarrow\infty} \frac{f_0^2}{TY}\frac{K^2}{4}\int_0^T \int_{-Y/2}^{Y/2} \Big(\mbox{e}^{-i\left(k(y+u)+\varphi(t+\tau)\right)}+c.c. \Big)\Big(\mbox{e}^{-i\left(ky+\varphi(t)\right)}+c.c. \Big) \,\mathrm{d}y\mathrm{d}t \nonumber\\ 
&=f_0^2+\lim_{Y,T\rightarrow\infty} \frac{f_0^2}{TY}\frac{K^2}{4}\Bigg(\mbox{e}^{-iku}\underbrace{\int_{-Y/2}^{Y/2} 1\,\mathrm{d}y}_{=Y}\int_0^T \mbox{e}^{-i\left(\varphi(t+\tau)-\varphi(t)\right)}\,\mathrm{d}t +c.c.+ \nonumber\\ 
&\qquad\qquad\qquad\qquad\quad+\mbox{e}^{-iku}\underbrace{\int_{-Y/2}^{Y/2} \mbox{e}^{-2iky}\,\mathrm{d}y}_{=\frac{\sin\left(kY\right)}{k}}\int_0^T \mbox{e}^{-i\left(\varphi(t+\tau)+\varphi(t)\right)}\,\mathrm{d}t +c.c. \Bigg)~.
\end{align}
In the limit $Y\gg \lambda$, again $\lim\limits_{Y\rightarrow\infty} \sin\left(kY\right)/kY\rightarrow 0$, yielding with equation (\ref{eq2})
\begin{align}\label{eq16}
\llangle f(y+u,&t+\tau) f(y,t)\rrangle_{y,t}\,=\\ \nonumber
&=f_0^2+\lim_{T\rightarrow\infty} \frac{f_0^2}{T}\frac{K^2}{4}\Bigg(\mbox{e}^{-iku} \int_0^T \mbox{e}^{-i\sum_{j=1}^N \varphi_j\cos\left(\omega_j (t+\tau) + \phi_j\right)}\mbox{e}^{i\sum_{j=1}^N \varphi_j\cos\left(\omega_j t + \phi_j\right)}\,\mathrm{d}t\, +\\ \nonumber
&\qquad\qquad\qquad\qquad+\mbox{e}^{iku} \int_0^T \mbox{e}^{i\sum_{j=1}^N \varphi_j\cos\left(\omega_j (t+\tau) + \phi_j\right)}\mbox{e}^{-i\sum_{j=1}^N \varphi_j\cos\left(\omega_j t + \phi_j\right)}\,\mathrm{d}t\Bigg)~.
\end{align}
Using equation (\ref{eq5}) for $N$ perturbation frequencies
\begin{equation}\label{eq17}
\mbox{e}^{\pm i\sum_{j=1}^N \varphi_j\cos\left(\omega_j t + \phi_j\right)}=\prod_{j=1}^N \sum_{n_j=-\infty}^{+\infty}\, \!J_{n_j}(\varphi_j)\mbox{e}^{in_j\left(\omega_j t+\phi_j\pm\frac{\pi}{2}\right)}~,
\end{equation}
equation (\ref{eq16}) can be rewritten, yielding
\begin{align}\label{eq18}
&\llangle f(y+u,t+\tau) f(y,t)\rrangle_{y,t}\,=\,f_0^2+\lim_{T\rightarrow\infty} \frac{f_0^2}{T}\frac{K^2}{4}\,\cdot \nonumber\\ 
&\qquad\cdot\Bigg(\mbox{e}^{-iku} \int_0^T\prod_{j=1}^N \sum_{n_j,m_j}\, \underbrace{\!J_{n_j}(\varphi_j)\,\!J_{m_j}(\varphi_j)\mbox{e}^{im_j\omega_j \tau}\mbox{e}^{i\left[n_j\left(\phi_j+\frac{\pi}{2}\right)+m_j\left(\phi_j-\frac{\pi}{2}\right)\right]}}_{=\,A^+_{n_j,m_j}}\cdot\mbox{e}^{i\left(n_j+m_j\right)\omega_j t}\,\mathrm{d}t\, + \nonumber\\ 
&\qquad+\mbox{e}^{iku} \int_0^T \prod_{j=1}^N \sum_{n_j,m_j}\, \underbrace{\!J_{n_j}(\varphi_j)\,\!J_{m_j}(\varphi_j)\mbox{e}^{im_j\omega_j \tau}\mbox{e}^{i\left[n_j\left(\phi_j-\frac{\pi}{2}\right)+m_j\left(\phi_j+\frac{\pi}{2}\right)\right]}}_{=\,A^-_{n_j,m_j}}\cdot\mbox{e}^{i\left(n_j+m_j\right)\omega_j t}\,\mathrm{d}t \Bigg) \nonumber\\ 
&=\,f_0^2+ \lim_{T\rightarrow\infty} \frac{f_0^2}{T}\frac{K^2}{4}\,\cdot \nonumber\\ 
&\qquad\cdot\Bigg(\mbox{e}^{-iku}\int_0^T \sum_{n_1,m_1}\cdots\sum_{n_N,m_N} A^+_{n_1,m_1}\cdots A^+_{n_N,m_N}\cdot \mbox{e}^{i\left(n_1+m_1\right)\omega_1 t}\cdots \mbox{e}^{i\left(n_N+m_N\right)\omega_N t}\,\mathrm{d}t\, + \nonumber\\ 
&\qquad+\mbox{e}^{iku}\int_0^T \sum_{n_1,m_1}\cdots\sum_{n_N,m_N} A^-_{n_1,m_1}\cdots A^-_{n_N,m_N}\cdot \mbox{e}^{i\left(n_1+m_1\right)\omega_1 t}\cdots \mbox{e}^{i\left(n_N+m_N\right)\omega_N t}\,\mathrm{d}t \Bigg) \nonumber\\ 
&=\,f_0^2+ f_0^2\frac{K^2}{4}\,\cdot \nonumber\\ 
&\qquad\cdot\Bigg(\mbox{e}^{-iku} \sum_{n_1,m_1}\cdots\sum_{n_N,m_N} A^+_{n_1,m_1}\cdots A^+_{n_N,m_N}\cdot\lim_{T\rightarrow\infty} \frac{1}{T}\int_0^T \mbox{e}^{i\sum_{j=1}^N\left(n_j+m_j\right)\omega_j t}\,\mathrm{d}t\, + \nonumber\\ 
&\qquad+\mbox{e}^{iku} \sum_{n_1,m_1}\cdots\sum_{n_N,m_N} A^-_{n_1,m_1}\cdots A^-_{n_N,m_N}\cdot\lim_{T\rightarrow\infty} \frac{1}{T}\int_0^T \mbox{e}^{i\sum_{j=1}^N\left(n_j+m_j\right)\omega_j t}\,\mathrm{d}t \Bigg)~.
\end{align}
A closer look to the time integral reveals, that it can only become zero or one
\begin{align}\label{eq19}
\lim_{T\rightarrow\infty} \frac{1}{T}\int_0^T \mbox{e}^{i\sum_{j=1}^N\left(n_j+m_j\right)\omega_j t}\,\mathrm{d}t=
\begin{cases} 1\quad \text{if}~~ \sum_{j=1}^N\left(n_j+m_j\right)\omega_j=0 \\ 0\quad \text{else} \end{cases}~.
\end{align}
This reduces the $2N$ sums in equation (\ref{eq18}) to a (single) sum over all integer multiplets $\left\{n_j, m_j\right\}\in\mathbb{Z}, j=1\ldots N$ for which the constraint
\begin{equation}\label{eq20}
\sum_{j=1}^N\left(n_j+m_j\right)\omega_j=0
\end{equation}
is fulfilled. Here it shall be noted, that for a finite acquisition time $T$, the constraint has to be changed to 
\begin{equation}\label{eq21}
\left|\sum_{j=1}^N\left(n_j+m_j\right)\omega_j\right|< 2\pi/T~,
\end{equation}
because the minimal resolvable frequency is defined by the measurement time via $1/T$. In the following, however, it is assumed that $T\rightarrow\infty$ and therefore equation (\ref{eq20}) is used for the calculations and discussions. Trivially, equation (\ref{eq20}) is satisfied for all integer multiplets with $n_j=-m_j$. However, depending on the specific values of $\omega_j$, the constraint might be fulfilled by additional integer multiplets $\left\{n_j, m_j\right\}$ with $n_j\neq -m_j$. The constraint from equation (\ref{eq20}) can be expressed mathematically by introducing a function $c$
\begin{align}\label{eq22}
c:\mathbb{Z}^N\times\mathbb{Z}^N\longrightarrow\mathbb{R}\quad,\quad\left(n_{j=1\ldots N},m_{j=1\ldots N}\right)\longmapsto \sum_{j=1}^N\left(n_j+m_j\right)\omega_j~,
\end{align}
with the kernel of $c$ $\left (\ker(c)\right )$ being the set of all integer multiplets $\left\{n_j, m_j\right\}\in\mathbb{Z} , j=1\ldots N$ for which $c\left( n_{j=1\ldots N},m_{j=1\ldots N} \right )=0$ and therefore the constraint in equation (\ref{eq20}) is fulfilled. Using this definition, equation (\ref{eq18}) simplifies to
\begin{align}\label{eq23}
\llangle f(y+u,t+\tau)&f(y,t)\rrangle_{y,t}\,=\,f_0^2+f_0^2\frac{K^2}{4}\,\cdot \nonumber\\ 
&\cdot\Bigg(\mbox{e}^{-iku}\sum_{\substack{\left\{n_j, m_j\right\}\in \ker(c)\\ j=1\ldots N}}\, \prod_{j=1}^N \, A^+_{n_j,m_j} +\mbox{e}^{iku}\sum_{\substack{\left\{n_j, m_j\right\}\in \ker(c)\\ j=1\ldots N}} \, \prod_{j=1}^N \, A^-_{n_j,m_j} \Bigg)~.
\end{align}

Using the result of equation (\ref{eq13}) and (\ref{eq23}), the second-order correlation function in equation (\ref{eq9}) now becomes
\begin{equation}
g^{(2)}(u,\tau) = 1 + \frac{K^2}{4}\left(\mbox{e}^{-iku}A_+ + \mbox{e}^{iku}A_-\right) ~,\label{eq24}
\end{equation}
with
\begin{align}
A_{\pm} &= \sum_{\substack{\left\{n_j, m_j\right\}\in \ker(c)\\ j=1\ldots N}}\, \prod_{j=1}^N \, \!J_{n_j}\left(\varphi_j\right)\!J_{m_j}\left(\varphi_j\right)\mbox{e}^{i m_j\omega_j\tau}\chi_{n_j,m_j}^\pm\label{eq25}\\
\chi_{n_j,m_j}^\pm &= \mbox{e}^{i\left[n_j\left(\phi_j\pm\frac{\pi}{2}\right)+m_j\left(\phi_j\mp\frac{\pi}{2}\right)\right]} ~. \label{eq26}
\end{align}

To calculate a more descriptive representation of the correlation function and to demonstrate, that $g^{(2)}(u,\tau)\in\mathbb{R}$, equation (\ref{eq24}), (\ref{eq25}) and (\ref{eq26}) can be further rewritten in terms of real and imaginary parts
\begin{align}\label{eq27}
g^{(2)}(u,\tau) = 1 + &\frac{K^2}{4}\sum_{\substack{\left\{n_j, m_j\right\}\in \ker(c)\\ j=1\ldots N}}\tilde{B}_{\{n_j, m_j\}}\left(\varphi_j\right)\mbox{e}^{i\left(\sum_{j=1}^N m_j\omega_j\tau+\sum_{j=1}^N \phi_j\left(m_j+n_j\right)\right)}\cdot \nonumber\\ 
&\cdot\left[\mbox{e}^{-i\left(ku+\frac{\pi}{2}\sum_{j=1}^N(m_j-n_j)\right)} +\mbox{e}^{i\left(ku+\frac{\pi}{2}\sum_{j=1}^N(m_j-n_j)\right)}\right] \nonumber\\ 
= 1 + &\frac{K^2}{2}\sum_{\substack{\left\{n_j, m_j\right\}\in \ker(c)\\ j=1\ldots N}}\tilde{B}_{\{n_j, m_j\}}\left(\varphi_j\right)\cos\left(ku+\tilde{\varphi}_{\{n_j, m_j\}}\right)\cdot \\ \nonumber
&\cdot\Biggl[\cos\Biggl(\sum_{j=1}^N m_j\omega_j\tau+\Phi_{\{n_j, m_j\}}\Biggr)+i\cdot\sin\Biggl(\sum_{j=1}^N m_j\omega_j\tau+\Phi_{\{n_j, m_j\}}\Biggr)\Biggr] ~,
\end{align}
with the product of the Bessel functions
\begin{equation} \label{eq28}
\tilde{B}_{\{n_j, m_j\}}\left(\varphi_j\right)\coloneqq \prod_{j=1}^N J_{n_j}\left(\varphi_j\right)J_{m_j}\left(\varphi_j\right) ~.
\end{equation}
The spatial correlation phase $\tilde{\varphi}_{\{n_j, m_j\}}$ and temporal phase $\Phi_{\{n_j, m_j\}}$ in equation (\ref{eq27}) are given by
\begin{align} \label{eq29}
\tilde{\varphi}_{\{n_j, m_j\}} &\coloneqq \frac{\pi}{2}\sum_{j=1}^N\left(m_j-n_j\right) \quad,\quad \Phi_{\{n_j, m_j\}}\coloneqq \sum_{j=1}^N \phi_j\left(m_j+n_j\right)~.
\end{align}
If the constraint in equation (\ref{eq20}) is fulfilled for a multiplet $\{n_j, m_j\}$, it is also satisfied for $-\{n_j, m_j\}$. With $\tilde{B}_{-\{n_j, m_j\}}\left(\varphi_j\right)=(-1)^{\sum_j(n_j+m_j)}\tilde{B}_{\{n_j, m_j\}}\left(\varphi_j\right)$, $\tilde{\varphi}_{-\{n_j, m_j\}}=-\tilde{\varphi}_{\{n_j, m_j\}}$ and $\Phi_{-\{n_j, m_j\}}=-\Phi_{\{n_j, m_j\}}$, it can be shown, that the addend with $-\{n_j, m_j\}$ of the sum in equation (\ref{eq27}) is complex conjugated to the addend with $\{n_j, m_j\}$. The addend of the zero multiplet $\{0,0\}$ is purely real valued. Therefore, the imaginary part vanishes after summing up all addends and equation (\ref{eq27}) becomes real
\begin{equation} \label{eq30}
g^{(2)}(u,\tau) = 1 + \frac{K^2}{2}\sum_{\substack{\left\{n_j, m_j\right\}\in \ker(c)\\ j=1\ldots N}}A_{\{n_j, m_j\}}\left(\tau,\Phi_{\{n_j, m_j\}}\right)\cos\left(ku+\tilde{\varphi}_{\{n_j, m_j\}}\right) ~,
\end{equation}
with the amplitude
\begin{equation} \label{eq31}
A_{\{n_j, m_j\}}\left(\tau,\Phi_{\{n_j, m_j\}}\right) =\tilde{B}_{\{n_j, m_j\}}\left(\varphi_j\right)\cdot\cos\Biggl(\sum_{j=1}^N m_j\omega_j\tau+\Phi_{\{n_j, m_j\}}\Biggr)~.
\end{equation}
Here, the sum has to be taken over all integer multiplets $\left\{n_j, m_j\right\}\in \ker(c), j=1\ldots N$ fulfilling the constraint in equation (\ref{eq20}). In principle, the constraint is satisfied for an infinite number of multiplets each with their own contribution to the correlation function given by 
\begin{equation} \label{eq32}
a_{\{n_j, m_j\}}(u,\tau) = A_{\{n_j, m_j\}}\left(\tau,\Phi_{\{n_j, m_j\}}\right)\cos\left(ku+\tilde{\varphi}_{\{n_j, m_j\}}\right) ~.
\end{equation}
However, contributions with large values of $\{n_j, m_j\}$ are suppressed, because the Bessel function (equation (\ref{eq28})) strongly decays for $m_j> \varphi_j$ \cite{abramowitz1964}. This limits the number of multiplets that have to be taken into account for the correlation analysis. The contribution to $g^{(2)}(u,\tau)$ of each multiplet addend $a_{\{n_j, m_j\}}(u,\tau)$ shows a periodic modulation in the correlation length $u$ with the same spatial periodicity $\lambda=2\pi/k$ as the unperturbed interference pattern, but shifted in $u$-direction by the spatial correlation phase $\tilde{\varphi}_{\{n_j, m_j\}}$. The amplitude $A_{\{n_j, m_j\}}\left(\tau,\Phi_{\{n_j, m_j\}}\right)$ (equation (\ref{eq31})) of each multiplet addend $a_{\{n_j, m_j\}}(u,\tau)$ depends on the correlation time $\tau$ and the specific perturbation characteristics. It shows a periodic structure in $\tau$ with the periodicity determined by the frequency component $\sum_{j=1}^N m_j\omega_j$ with the coefficients $\{n_j, m_j\}\in\ker(c)$. This periodic structure is shifted in $\tau$-direction by the temporal phase $\Phi_{\{n_j, m_j\}}$. The amplitude of the modulation in $\tau$-direction is given by the peak phase deviations $\varphi_j$, via the product of the Bessel functions in equation (\ref{eq28}). 

After summing up all addends $a_{\{n_j, m_j\}}(u,\tau)$, the resulting correlation function (equation (\ref{eq30})) shows the same spatial periodicity $\lambda$. The overall amplitude is equal to 1 only at certain correlation times $\tau$ given by the involved perturbation frequencies (see section \ref{sec2.4}). At these temporal positions, the contrast of the correlation function is $K^2/2$ and therefore directly linked to the contrast of the unperturbed interference pattern $K$. For all other correlation times, the contrast of the correlation function is $<K^2/2$. It has to be noted, that the maximum contrast of the correlation function is $K^2/2$ and therefore a factor of $K/2$ lower than the contrast $K$ of the unperturbed interference pattern. 

The overall amplitude of the resulting correlation function includes the perturbation frequencies $\omega_j$, their harmonic frequencies as well as their differences and sums (intermodulation terms). All frequency components are given by the argument of the cosine $\sum_{j=1}^N m_j\omega_j$ in equation (\ref{eq31}). Approximately, the maximum frequency component per perturbation frequency included in the correlation function is given by $m_{j,\text{max}}\omega_j$, with $m_{j,\text{max}}\approx\varphi_j$, as larger frequency components are suppressed due to the strong decay of the Bessel function in equation (\ref{eq28}) for $m_j> \varphi_j$. Therefore, the maximum frequency component of all perturbation frequencies included in the correlation function is approximately given by $\text{max}\{\varphi_j\omega_j\}$.

\subsection{Approximate solution for the second-order correlation function}\label{sec2.3}
In the following, an approximate solution for the correlation function is deduced, by taking into account only the trivial solution to the constraint in equation (\ref{eq20}). These are the multiplets with $n_j=-m_j$, which typically give the main contribution to the correlation function. Using $\Phi_{\{-m_j, m_j\}}=0$, $\tilde{\varphi}_{\{-m_j,m_j\}} =\pi\sum_{j=1}^N m_j$ and $\!J_{-m_j}(\varphi_j)=\left(-1\right)^{m_j}\!J_{m_j}(\varphi_j)$, equation (\ref{eq30}) and (\ref{eq31}) become
\begin{align} \label{eq33}
&g^{(2)}(u,\tau) = 1 + \frac{K^2}{2}\cdot \\ \nonumber
&\cdot\sum_{\substack{\left\{-m_j,m_j\right\}\in \ker(c)\\ j=1\ldots N}}\,\left(\prod_{j=1}^N J_{m_j}\left(\varphi_j\right)^2\right)\cos\Biggl(\sum_{j=1}^N m_j\omega_j\tau\Biggr)\underbrace{\left(-1\right)^{\sum_{j=1}^N m_j}\cos\left(ku+\pi\sum_{j=1}^N m_j\right)}_{=\,\cos\left(ku \right)} ~.
\end{align}
With $\cos\left(\sum_{j=1}^N m_j\omega_j\tau\right)=\frac{1}{2}\left(\mbox{e}^{i\sum_{j=1}^N m_j\omega_j\tau}+\mbox{e}^{-i\sum_{j=1}^N m_j\omega_j\tau}\right)$ and $\mbox{e}^{i\sum_{j=1}^N m_j\omega_j\tau}=\prod_{j=1}^N \mbox{e}^{i m_j\omega_j\tau}$, the approximate second-order correlation function yields
\begin{equation}\label{eq34}
g^{(2)}(u,\tau) = 1 + \frac{K^2}{2}\cdot A\left(\tau\right)\cos\left(ku\right) ~,
\end{equation}
with the time-dependent amplitude
\begin{align}\label{eq35}
A(\tau) &=\prod_{j=1}^N\Biggl(\sum_{m_j=-\infty}^\infty J_{m_j}\left(\varphi_j\right)^2\cos\left(m_j\omega_j\tau\right)\Biggr) \\ \nonumber
&= \prod_{j=1}^N \Biggl(J_{0}\left(\varphi_j\right)^2 +2\cdot\sum_{m_j=1}^\infty J_{m_j}\left(\varphi_j\right)^2\cos\left(m_j\omega_j\tau\right)\Biggr)~.
\end{align}
Similar to the explicit solution of the correlation function in equation (\ref{eq30}) and (\ref{eq31}), the spatial modulation of the approximate solution is given by the spatial periodicity $\lambda$ of the unperturbed interference pattern. The approximate solution is independent of the spatial correlation phases $\tilde{\varphi}_{\{n_j, m_j\}}$ and temporal phases $\Phi_{\{n_j, m_j\}}$ (equation (\ref{eq29})). Therefore, the addends are not phase shifted with respect to each other in $u$- and $\tau$-direction. 

Usually, the approximate correlation function in equation (\ref{eq34}) can be used for the description of multifrequency perturbations. However, in the case of few perturbation frequencies, that are multiples of each other, the constraint in equation (\ref{eq20}) is additionally fulfilled for $n_j\neq -m_j$ and the explicit solution of the correlation function in equation (\ref{eq30}) has to be applied. In the case of a single perturbation frequency, the constraint of equation (\ref{eq20}) is only satisfied for $n_1= -m_1$. Thus, the explicit and approximate solution of the correlation function are identical.

\subsection{Determination of contrast and spatial periodicity}\label{sec2.4}
The determination of the contrast $K$ and spatial periodicity $\lambda$ from the correlation function shall be discussed now. Both can only be correctly obtained at certain correlation times. As mentioned in section \ref{sec2.2} and \ref{sec2.3}, the explicit and approximate solution of the correlation function show a periodic modulation in $u$-direction having the same periodicity $\lambda$ as the unperturbed interference pattern. The overall amplitude of this modulation (after summing up all multiplet addends) depends on the correlation time $\tau$ resulting from the specific perturbation spectrum for both correlation functions (equation (\ref{eq30}) and (\ref{eq34})). Its maximum value of 1 is achieved at $\tau = M_\tau\tau_s, M_\tau\in\mathbb{N}_0$, with the superperiod $\tau_s=2\pi/\omega_{gcd}$ given by the reciprocal value of the greatest common divisor of all perturbation frequencies $gcd(\omega_1, \ldots, \omega_N)=\omega_{gcd}$. For each frequency $\omega_j$ there is an integer $s_j\in\mathbb{N}$ for which $\omega_{gcd}=\omega_j/s_j$. At the temporal positions $\tau = M_\tau\tau_s$, only the addends with $n_j=-m_j$ sum up to a maximum value of 1, because then the addends in equation (\ref{eq30}) are not phase shifted in $u$- and $\tau$-direction with respect to each other. The sum of the addends with $n_j\neq -m_j$ is equal to zero at these temporal positions. Therefore, the exact and approximate solution are identical at correlation times $\tau=M_\tau\tau_s$. Using equation (\ref{eq34}) and (\ref{eq35}), the correlation function then becomes
\begin{align}\label{eq36}
g^{(2)}(u,M_\tau\tau_s) &= 1 + \frac{K^2}{2}\cos\left(ku\right)\cdot \prod_{j=1}^N \underbrace{\Biggl(\sum_{m_j=-\infty}^\infty J_{m_j}\left(\varphi_j\right)^2\underbrace{\cos\left(2\pi M_\tau m_j s_j\right)}_{=\,1}\Biggr)}_{=\,1}\nonumber\\ 
&=1 + \frac{K^2}{2}\cos\left(ku\right) ~,
\end{align}
which is suitable to obtain the contrast $K$ and pattern periodicity $\lambda=2\pi/k$ of the unperturbed interference pattern. If there is no greatest common divisor $\omega_{gcd}$, the superperiod $\tau_s$ is infinite and the only position, where the amplitudes have a maximum value of $1$, is at $\tau=0$. Therefore, the determination of the contrast and spatial periodicity using equation (\ref{eq36}) can always be applied to the correlation function at the correlation time of $\tau=0$.

\subsection{Correlation function of single- and two-frequency perturbations}\label{sec2.5}
The properties of the explicit and approximate correlation function, discussed in section \ref{sec2.2} and \ref{sec2.3}, are illustrated below for single- and two-frequency perturbations. The commonalities and differences of both solutions are pointed out.

\begin{figure}
\centering
\includegraphics[width=0.5\textwidth]{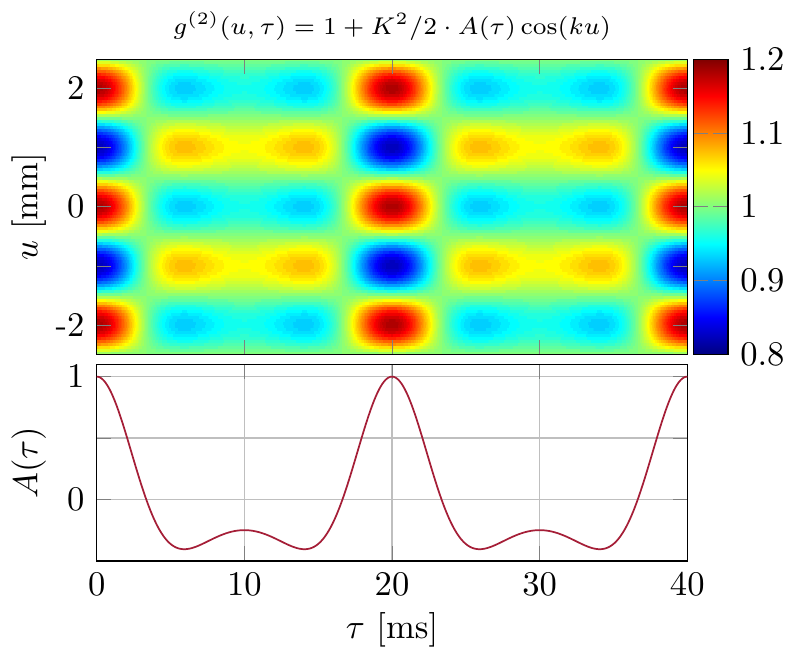} \caption{Top: Correlation function of an interference pattern with $K=0.6$, $\lambda=\unit[2]{mm}$ and a perturbation consisting of $\omega_1/2\pi = \unit[50]{Hz}$, $\varphi_1=\unit[0.76]{\pi}$ calculated with equation (\ref{eq34}) and (\ref{eq35}) for $N=1$. Bottom: Correlation amplitude depending on the correlation time $\tau$ and the perturbation parameters resulting from equation (\ref{eq35}). It has its maximum value of 1 at the temporal coordinates $\tau=M_\tau\cdot2\pi/\omega_1=M_\tau\cdot\unit[20]{ms}$, with $M_\tau\in\mathbb{N}_0$.}
\label{fig2}
\end{figure}

In the case of one perturbation frequency $\omega_1$ ($N=1$), the constraint in equation (\ref{eq20}) is only fulfilled for $n_1=-m_1$. Thus, the explicit and approximate solution are identical. A correlation function of an interference pattern with $K=0.6$, $\lambda=\unit[2]{mm}$, a perturbation frequency $\omega_1/2\pi = \unit[50]{Hz}$ and a peak phase deviation $\varphi_1=\unit[0.76]{\pi}$ is calculated according to equation (\ref{eq34}) and (\ref{eq35}) for $N=1$ and plotted in figure \ref{fig2} at the top, showing a clear periodic structure in the correlation length $u$ and time $\tau$. As seen from equation (\ref{eq34}), the spatial periodicity of the unperturbed interference pattern is recovered in the correlation function as modulation along $u$. At each $\tau$, however, the contrast is reduced to $K^2/2\cdot A(\tau)$. As seen from equation (\ref{eq35}) and figure \ref{fig2} at the bottom, the amplitude $A(\tau)$ reaches its maximum at correlation times that are multiples of the perturbation time period $2\pi/\omega_1=\unit[20]{ms}$. Here, $A(\tau)$ becomes 1 and the contrast in the correlation function $K^2/2$ is directly linked to the contrast of the unperturbed interference pattern. According to equation (\ref{eq7}) and figure \ref{fig1}, the integrated interference pattern would be completely ``washed-out" for these perturbation parameters. Using correlation theory, however, the contrast and spatial periodicity can be unveiled as described in section \ref{sec2.4}.

\begin{figure}
\centering
\includegraphics[width=1\textwidth]{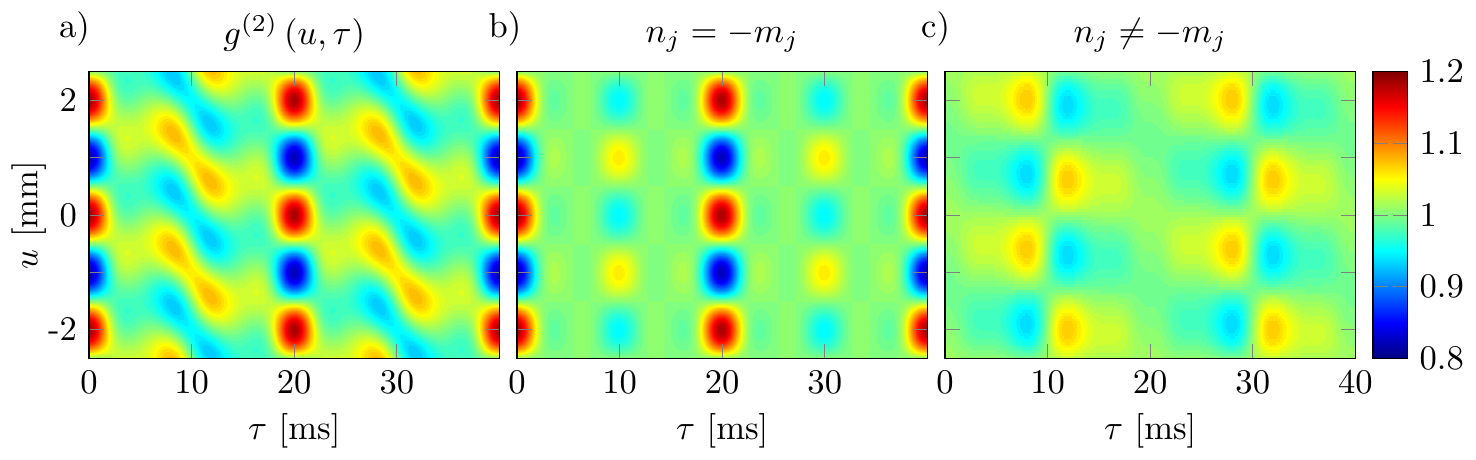} \caption{a) Second-order correlation function $g^{(2)}(u,\tau)$ of an interference pattern with a contrast of $K=0.6$, a spatial periodicity $\lambda=\unit[2]{mm}$ perturbed with two frequencies $\omega_1/2\pi=\unit[50]{Hz}, \varphi_1=\unit[0.5]{\pi}, \phi_1=\unit[0.25]{\pi}$ and $\omega_2/2\pi=\unit[100]{Hz}, \varphi_2=\unit[0.5]{\pi}, \phi_2=\unit[-0.25]{\pi}$ calculated according to equation (\ref{eq30}) and (\ref{eq31}). b) Approximate solution calculated with equation (\ref{eq34}) and (\ref{eq35}) including only the $n_j=-m_j$ terms. c) Difference between explicit and approximate solution solely given by the terms with $n_j\neq -m_j$.}
\label{fig3}
\end{figure}

In the case of perturbation frequencies, that are multiples of each other, the explicit solution of the correlation function has to be used. This shall be demonstrated for two frequencies with $\omega_2=2\omega_1$. Here, equation (\ref{eq20}) becomes $(n_1+m_1)=-2(n_2+m_2)$ and the constraint is satisfied not only for integer multiplets with $n_{1/2}=-m_{1/2}$, but also for $n_{1/2}\neq -m_{1/2}$, as $\{n_1,m_1,n_2,m_2\}=\{1,1,-1,0\}$. These terms lead to additional contributions to the explicit correlation function in equation (\ref{eq30}), causing a spatial and temporal phase shift due to the not vanishing phases $\tilde{\varphi}_{\{n_j, m_j\}}$ and $\Phi_{\{n_j, m_j\}}$ (equation (\ref{eq29})). Therefore, the approximate solution of equation (\ref{eq34}) is not suitable, and the explicit solution has to be used. In figure \ref{fig3}(a), a second-order correlation function calculated with the explicit solution (equation (\ref{eq30}) and (\ref{eq31})) is shown for an interference pattern with a contrast of $K=0.6$ and a spatial periodicity $\lambda=\unit[2]{mm}$ perturbed with $\omega_1/2\pi=\unit[50]{Hz}, \varphi_1=\unit[0.5]{\pi}, \phi_1=\unit[0.25]{\pi}$ and $\omega_2/2\pi=\unit[100]{Hz}, \varphi_2=\unit[0.5]{\pi}, \phi_2=\unit[-0.25]{\pi}$. The approximate solution (equation (\ref{eq34}) and (\ref{eq35})) including only the $n_j=-m_j$ terms can be seen in figure \ref{fig3}(b) and differs clearly from the exact solution in (a). Both correlation functions, figure \ref{fig3}(a) and (b), show a periodic modulation in $u$-direction having the same periodicity $\lambda$ as the unperturbed interference pattern. The superperiod $\tau_s=2\pi/\omega_{gcd}=\unit[20]{ms}$ can be identified in both correlation functions originating from the greatest common divisor of the involved perturbation frequencies $\omega_{gcd}/2\pi=\unit[50]{Hz}$. At correlation times, which are multiples of this superperiod, the explicit and approximate correlation function are identical and the amplitudes achieve their maximum value of 1 (see section \ref{sec2.4}). The difference between the exact and approximate solution, which is solely given due to the terms with $n_j\neq -m_j$, is illustrated in figure \ref{fig3}(c). Its contribution to the correlation function in figure \ref{fig3}(a) leads to the correlation time dependent phase shift of the pattern in $u$-direction originating from the spatial correlation phases $\tilde{\varphi}_{\{n_j, m_j\}}$ and temporal phases $\Phi_{\{n_j, m_j\}}$.
\begin{figure}
\centering
\includegraphics[width=1\textwidth]{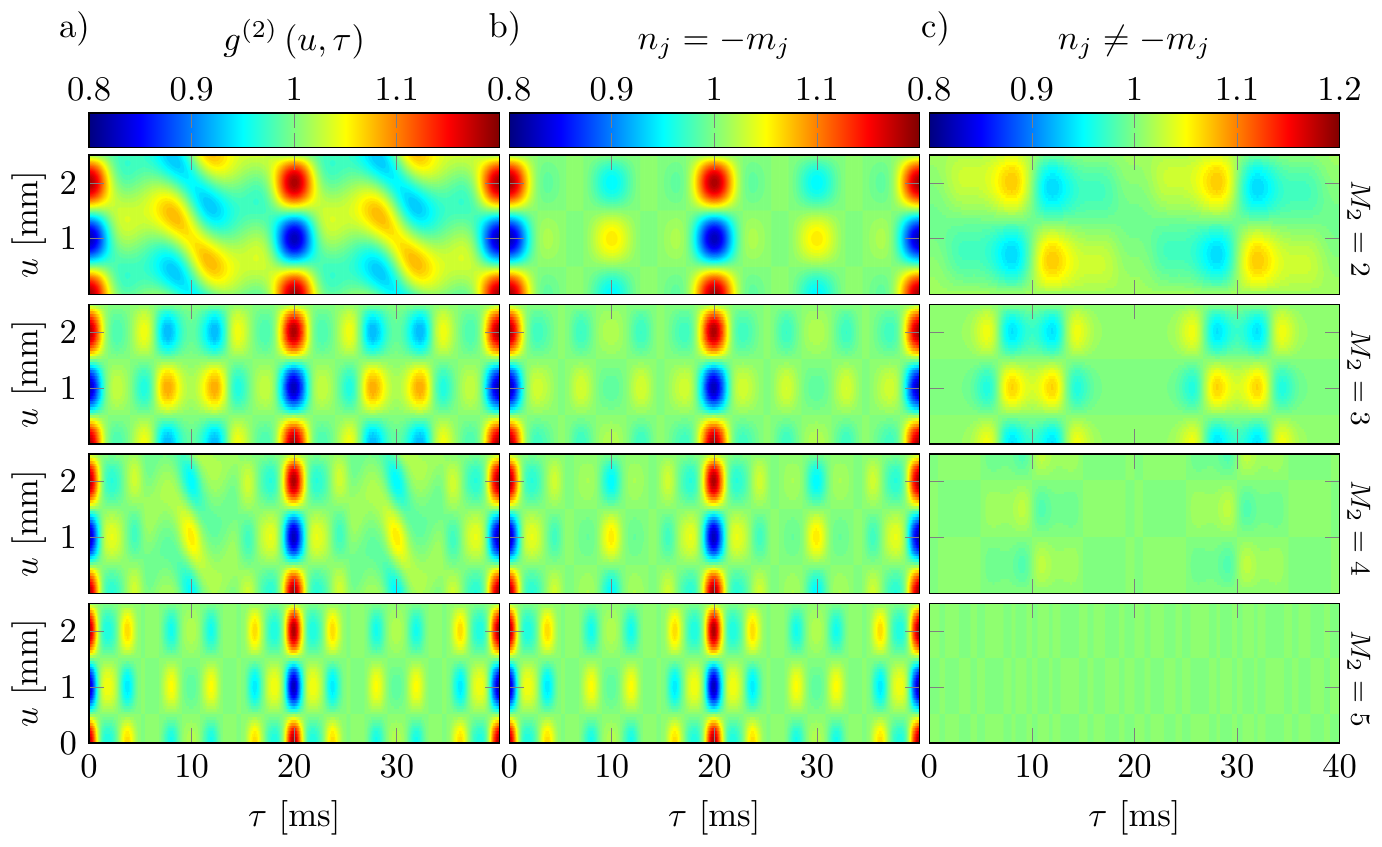} \caption{a) Shown is a perturbation consisting of two frequencies with $\omega_1/2\pi=\unit[50]{Hz}, \varphi_1=\unit[0.5]{\pi}, \phi_1=\unit[0.25]{\pi}$ and $\omega_2=M_2\omega_1, M_2\in\{2,3,4,5\}$ with $\varphi_2=\unit[0.5]{\pi}, \phi_2=\unit[-0.25]{\pi}$ as calculated according to equation (\ref{eq30}). b) Approximate solution of the second-order correlation function calculated with equation (\ref{eq34}). c) Difference between explicit and approximate solution given by the terms with $n_j\neq -m_j$. The structure in c) is almost vanished for $M_2=4$ and disappeared for $M_2=5$. At this transition point, the correlation functions in a) and b) are identical and the approximate solution is suitable for the analysis.}
\label{fig4}
\end{figure}

\subsection{Transition between explicit and approximate solution}\label{sec2.6}
As discussed in section \ref{sec2.3}, the explicit and approximate solution (equation (\ref{eq30}) and (\ref{eq34})) are identical in the case of one perturbation frequency ($N=1$) and also in the case of numerous frequencies $\omega_j$, that are not multiples of each other. A common scenario, when both solutions differ from each other, is the case when the frequencies are multiples of each other, e.g. for $N$ perturbation frequencies $\omega_j=M_j\omega_1$ with $j=2\ldots N, M_j\in \mathbb{N}\backslash\{0,1\}$ and $M_2<\ldots<M_N$. Here, it shall be shown, that also in this case the approximate solution might be suitable to describe the correlation function.

The explicit solution turns into the approximate, if the largest contributing addend in equation (\ref{eq30}) for $n_j\neq -m_j$ is small compared to the largest addend with $n_j=-m_j$. The amplitudes of all addends are given by the peak phase deviations $\varphi_j$, via the product of the Bessel functions in equation (\ref{eq28}). The maximum value of the Bessel function $J_m\left(\varphi\right)$ is approximately achieved, if the order is $m\approx\varphi-1$ for $\varphi\geq 1$ and $m=0$ otherwise. Therefore, the amplitude of the largest addend with $n_j=-m_j$ is given by
\begin{equation}\label{eq37}
\tilde{B}_{\{n_j= -m_j\}}(\varphi_j)\approx \prod_{j=1}^N J_{\varphi_j-1}\left(\varphi_j\right)^2~.
\end{equation}
Using equation (\ref{eq20}), the constraint for the integer multiplets $\left\{n_j, m_j\right\}\in \ker(c)$ can be written as
\begin{equation}\label{eq38}
\left(n_1+m_1\right)=-\sum_{j=2}^N M_j\left(n_j+m_j\right)~.
\end{equation}
This constraint is generally satisfied for the multiplet with $n_j\neq -m_j$, resulting in the largest contributing integer multiplet
\begin{align}\label{eq39}
\big\{n_1,m_1,n_2,m_2,n_3,m_3,&\ldots,n_N,m_N\big\}= \nonumber \\
&=\left\{\frac{M_2}{2},\frac{M_2}{2},-(m_2+1),m_2,-m_3,m_3,\ldots,-m_N,m_N\right\}~.
\end{align} 
Using this multiplet and $m\approx\varphi-1$ for the order of the Bessel function, the amplitude (equation (\ref{eq28})) of the largest addend with $n_j\neq -m_j$ contributing to the correlation function is
\begin{equation}\label{eq40}
\tilde{B}_{\{n_j\neq -m_j\}}(\varphi_j)\approx J_{\frac{M_2}{2}}\left(\varphi_1\right)^2 J_{-\varphi_2}\left(\varphi_2\right)J_{\varphi_2-1}\left(\varphi_2\right)\prod_{j=3}^N J_{\varphi_j-1}\left(\varphi_j\right)^2~.
\end{equation}
Here, the variable factor $J_{\frac{M_2}{2}}\left(\varphi_1\right)^2$ is given by $M_2=\omega_2/\omega_1$ and determines the value of $\tilde{B}_{\{n_j\neq -m_j\}}(\varphi_j)$. The ratio of equation (\ref{eq37}) and (\ref{eq40}) is 
\begin{equation}\label{eq41}
\frac{\tilde{B}_{\{n_j\neq -m_j\}}(\varphi_j)}{\tilde{B}_{\{n_j= -m_j\}}(\varphi_j)}=\frac{J_{\frac{M_2}{2}}\left(\varphi_1\right)^2 J_{-\varphi_2}\left(\varphi_2\right)}{J_{\varphi_1-1}\left(\varphi_1\right)^2 J_{\varphi_2-1}\left(\varphi_2\right)}~.
\end{equation} 
A closer look to the Bessel function unveils that $J_{\frac{M_2}{2}}\left(\varphi_1\right)^2$ decays rapidly for $M_2/2>\varphi_1$, which can be seen with the asymptotic form of the Bessel function for $0<\varphi_1\ll\sqrt{n+1}$ \cite{abramowitz1964}
\begin{equation}\label{eq42}
J_n\left(\varphi_1\right)\approx \frac{1}{\Gamma \left(n+1\right)}\left(\frac{\varphi_1}{2}\right)^n~,
\end{equation}
with $\Gamma \left(n+1\right)=n!$ denoting the gamma function. Therefore, the explicit solution approaches rapidly to the approximate solution, once $M_2>2\varphi_1$.

Figure \ref{fig4} shows an example for a two frequency perturbation with $M_2=\omega_2/\omega_1=\{2,3,4,5\}$. For $\varphi_1=\unit[0.5]{\pi}$, the explicit solution should closely approach the approximate once $M_2>2\varphi_1\approx 3$, as can be seen in figure \ref{fig4}. For $M_2=4$, the structure in figure \ref{fig4}(c) is almost vanished. In this case, the leading term of equation (\ref{eq40}) is about seven times smaller than $\tilde{B}_{\{n_j= -m_j\}}(\varphi_j)$ in equation (\ref{eq37}). The structure in figure \ref{fig4}(c) disappears for $M_2=5$. Here, $\tilde{B}_{\{n_j\neq -m_j\}}(\varphi_j)$ is twenty-four times smaller than $\tilde{B}_{\{n_j= -m_j\}}(\varphi_j)$ and has a negligible contribution to the correlation function. Therefore, the correlation functions in figure \ref{fig4}(a) and (b) for $M_2=5$ are identical, proving the approximate solution to be suitable for the correlation analysis.

\subsection{Amplitude spectrum of the second-order correlation function}\label{sec2.7}
As it might be difficult to identify the involved perturbation frequencies from the correlation function (c.f. figure \ref{fig3}), the amplitude spectrum of the correlation function can be calculated and used for the determination of the perturbation characteristics \cite{Rembold2016}. This is possible, as the Fourier transform of the correlation function equals the power spectrum of the perturbed measurement signal according to the Wiener-Khintchine theorem \cite{Wiener1930,Khintchine1934}. Therefore, the applied perturbation frequencies can be identified in the amplitude spectrum of $g^{(2)}(u,\tau)$. To determine the frequency components and their amplitudes, the temporal Fourier transform $\mathcal{F}\left(g^{(2)}(u,\tau)\right)(u,\omega)=\frac{1}{\sqrt{2\pi}}\int_{-\infty}^\infty g^{(2)}(u,\tau)\mbox{e}^{i\omega\tau}\,\mathrm{d}\tau$ of equation (\ref{eq30}) is calculated
\begin{align} \label{eq43}
\mathcal{F}\Big(g^{(2)}(u,\tau)\Big)(u,\omega) &=\, \sqrt{2\pi}\,\delta(\omega) + \frac{K^2}{2}\cdot\\ \nonumber
&\cdot\sum\limits_{\substack{\left\{n_j, m_j\right\}\in \ker(c)\\ j=1\ldots N}}\mathcal{F}\Big(A_{\{n_j, m_j\}}\left(\tau,\Phi_{\{n_j, m_j\}}\right)\Big)(\omega)\cdot\cos\left(ku+\tilde{\varphi}_{\{n_j, m_j\}}\right)~, \nonumber
\end{align}
with the Fourier transformed amplitude of equation (\ref{eq31})
\begin{align} \label{eq44}
&\mathcal{F}\Big(A_{\{n_j, m_j\}}\left(\tau,\Phi_{\{n_j, m_j\}}\right)\Big)(\omega)=\\ \nonumber
&= \frac{\tilde{B}_{\{n_j, m_j\}}\left(\varphi_j\right)}{2\sqrt{2\pi}}\Bigg(\mbox{e}^{-i\Phi_{\{n_j, m_j\}}}\int_{-\infty}^\infty\mbox{e}^{i\left(\omega-\sum\limits_{j=1}^{N}m_j\omega_j\right)\tau}\,\mathrm{d}\tau\,+\mbox{e}^{i\Phi_{\{n_j, m_j\}}}\int_{-\infty}^\infty\mbox{e}^{i\left(\omega+\sum\limits_{j=1}^{N}m_j\omega_j\right)\tau}\,\mathrm{d}\tau\Bigg) \\ \nonumber
&= \frac{\sqrt{2\pi}\,\tilde{B}_{\{n_j, m_j\}}\left(\varphi_j\right)}{2}\Bigg[\mbox{e}^{-i\Phi_{\{n_j, m_j\}}}\delta \left(\omega-\sum\limits_{j=1}^{N}m_j\omega_j\right)+\mbox{e}^{i\Phi_{\{n_j, m_j\}}}\delta \left(\omega+\sum\limits_{j=1}^{N}m_j\omega_j\right)\Bigg]~.
\end{align}
Using $\mbox{e}^{\pm i\Phi_{\{n_j, m_j\}}}=\cos\left(\Phi_{\{n_j, m_j\}}\right)\pm i\cdot\sin\left(\Phi_{\{n_j, m_j\}}\right)$ and $\omega_{\{m_j\}} \coloneqq \sum\limits_{j=1}^{N}m_j\omega_j$, the amplitude in equation (\ref{eq44}) results in
\begin{align} \label{eq45}
\mathcal{F}\Big(A_{\{n_j, m_j\}}\left(\tau,\Phi_{\{n_j, m_j\}}\right)\Big)&(\omega) = \sqrt{2\pi}\,\tilde{B}_{\{n_j, m_j\}}\left(\varphi_j\right)\cdot\\ \nonumber
&\cdot\bigg(\cos\left(\Phi_{\{n_j, m_j\}}\right)\Big(\delta(\omega+\omega_{\{m_j\}}) + \delta(\omega-\omega_{\{m_j\}})\Big)+\\ \nonumber
&+i\cdot\sin\left(\Phi_{\{n_j, m_j\}}\right)\Big(\delta(\omega+\omega_{\{m_j\}}) - \delta(\omega-\omega_{\{m_j\}})\Big)\bigg) ~,
\end{align}
where $\omega_{\{m_j\}}$ are the frequency components and $\delta\left(\omega\right)$ the Dirac delta function. The resulting amplitude spectrum $\left|\mathcal{F}\left(g^{(2)}(u,\tau)\right)(u,\omega)\right|$ calculated from equation (\ref{eq43}) and (\ref{eq45}) is given by
\begin{align} \label{eq46}
\frac{1}{2\pi}\Big|\mathcal{F}&\left(g^{(2)}(u,\tau)\right)(u,\omega)\Big|^2 = \delta(\omega)^2\,+ \\ \nonumber
&+ 2\cdot\Bigg(\frac{K^2}{2}\sum\limits_{\substack{\left\{n_j, m_j\right\}\in\ker(c)\\ j=1\ldots N}} \tilde{B}_{\{n_j, m_j\}}\left(\varphi_j\right)\cdot\cos\left(ku+\tilde{\varphi}_{\{n_j, m_j\}}\right)\cdot\\ \nonumber
&\cdot\bigg(\cos\left(\Phi_{\{n_j, m_j\}}-\frac{\pi}{4}\right)\delta(\omega+\omega_{\{m_j\}})+\cos\left(\Phi_{\{n_j, m_j\}}+\frac{\pi}{4}\right)\delta(\omega-\omega_{\{m_j\}})\bigg)\Bigg)^2 ~. \nonumber
\end{align}
As for the explicit correlation function in equation (\ref{eq30}), the sum in equation (\ref{eq46}) is taken over the integer multiplets $\left\{n_j, m_j\right\}\in \ker(c), j=1\ldots N$ resulting from the constraint in equation (\ref{eq20}). Each addend shows a periodic structure in the correlation length $u$ with the same spatial periodicity $\lambda$ as the unperturbed interference pattern, but shifted in $u$-direction by the spatial correlation phase $\tilde{\varphi}_{\{n_j, m_j\}}$ (equation (\ref{eq29})). The amplitude of this modulation depends on the peak phase deviation $\varphi_j$, via the product of the Bessel functions in equation (\ref{eq28}) modified by the cosine of the temporal phase $\Phi_{\{n_j, m_j\}}$ (equation (\ref{eq29})). The frequency position of the modulation is given by the frequency component $\omega_{\{m_j\}} \coloneqq \sum_{j=1}^N m_j\omega_j$ with the coefficients $\{n_j,m_j\}\in\ker(c)$. The summed up amplitude spectrum contains all frequency components up to a maximum frequency $\text{max}\{\varphi_j\omega_j\}$ resulting from the strong decay of the Bessel function as discussed for the explicit solution of the correlation function (equation (\ref{eq30})).
\begin{figure}
\centering
\includegraphics[width=1\textwidth]{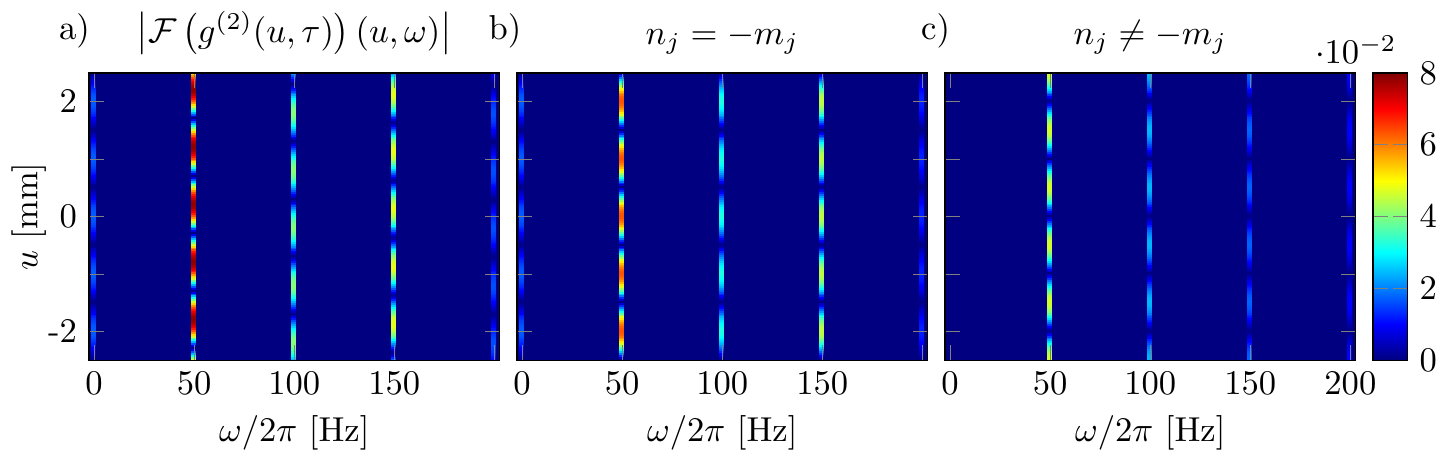} \caption{a) Amplitude spectrum $\Big|\mathcal{F}\left(g^{(2)}(u,\tau)\right)(u,\omega)\Big|$ of a perturbation with $\omega_1/2\pi=\unit[50]{Hz}, \varphi_1=\unit[0.5]{\pi}, \phi_1=\unit[0.25]{\pi}$ and $\omega_2/2\pi=\unit[100]{Hz}, \varphi_2=\unit[0.5]{\pi}, \phi_2=\unit[-0.25]{\pi}$ calculated according to equation (\ref{eq46}). b) Approximate solution of the amplitude spectrum calculated using equation (\ref{eq47}). c) Contribution of the addends with $n_j\neq -m_j$ to the amplitude spectrum in a).}
\label{fig5}
\end{figure}

The amplitude spectrum of the approximate second-order correlation function in equation (\ref{eq34}) is calculated from equation (\ref{eq46}) using $n_j=-m_j$ and reads
\begin{align}\label{eq47}
\frac{1}{2\pi}&\Big|\mathcal{F}\left(g^{(2)}(u,\tau)\right)(u,\omega)\Big|^2 = \delta(\omega)^2+ \\ \nonumber
&+ \Bigg(\frac{K^2}{2}\cos\left(ku\right)\sum\limits_{\substack{\left\{-m_j,m_j\right\}\in\ker(c)\\ j=1\ldots N}} \,\left(\prod\limits_{j=1}^N J_{m_j}(\varphi_j)^2\right)\cdot\bigg(\delta(\omega+\omega_{\{m_j\}})+\delta(\omega-\omega_{\{m_j\}})\bigg)\Bigg)^2 ~. \nonumber
\end{align}
As with the approximate solution, also the amplitude spectrum is independent from the spatial correlation phases $\tilde{\varphi}_{\{n_j, m_j\}}$ and temporal phases $\Phi_{\{n_j, m_j\}}$. Therefore, the spatial periodicity in the amplitude spectrum is directly given by the periodicity of the unperturbed interference pattern. The amplitude of the modulation in $u$-direction is now solely given by the peak phase deviations $\varphi_j$, via the square of the Bessel functions. 

The amplitude spectrum corresponding to the second-order correlation function of the two frequency perturbation shown in figure \ref{fig3}(a) is calculated according to equation (\ref{eq46}) and plotted in figure \ref{fig5}(a) for positive frequency components. The amplitude spectrum of the approximate solution of equation (\ref{eq47}) is plotted in figure \ref{fig5}(b). In figure \ref{fig5}(c), the contribution of the addends with $n_j\neq -m_j$ to the explicit solution is shown. For each frequency component, given by $m_1\omega_1+m_2\omega_2$ with $\{n_1,m_1,n_2,m_2\}\in \ker(c)$, the modulation of the unperturbed interference pattern with the periodicity $\lambda$ can be identified in the spatial direction $u$. In figure \ref{fig5}(c), the phase shift of the interference pattern in the spatial direction is caused by the spatial phases $\tilde{\varphi}_{\{n_j, m_j\}}$. In figure \ref{fig5}(b), the interference pattern is not shifted in the spatial direction $u$ and the amplitudes of the involved frequency components only depend on the peak phase deviations $\varphi_j$ via the square of the Bessel functions in equation (\ref{eq47}). As for the transition between the explicit and approximate correlation function discussed in section \ref{sec2.6}, the amplitude spectrum of the explicit correlation function (equation (\ref{eq46})) turns into the approximate solution (equation (\ref{eq47})), if the largest contributing addend in equation (\ref{eq40}) is small compared to equation (\ref{eq37}).

\subsection{Determination of perturbation characteristics}\label{sec2.8}
The identification of unknown perturbation frequencies from the correlation function is difficult, because the structure of the correlation function does not reveal the frequencies directly (c.f. figure \ref{fig3}). The amplitude spectrum offers a better access to the frequency components included in the correlation function. In \cite{Rembold2016}, a numerical algorithm is described, which allows to identify unknown perturbation frequencies. After the determination of the perturbation frequencies $\omega_j$, equation (\ref{eq46}) or (\ref{eq47}) is applied to obtain the corresponding peak phase deviations $\varphi_j$ and phases $\phi_j$. The latter are not uniquely determined, as the correlation function is invariant under phase transformations. However, it is possible to identify the original phases by the reconstruction of the unperturbed interference pattern with the obtained perturbation frequencies and amplitudes.

\subsubsection{Invariance of the correlation function}\label{sec2.8.1}
As discussed in section \ref{sec2.4}, the contrast and pattern periodicity of the unperturbed interference pattern can be unveiled from the second-order correlation function. Using the amplitude spectrum, also the perturbation frequencies, amplitudes and phases can be determined (section \ref{sec2.7}). However, the perturbation phases can not be uniquely obtained. For a single perturbation frequency, for instance, the phase information is completely lost in the correlation function. However, for larger numbers of perturbation frequencies ($N>1$), the phase informations are not completely lost. For $N>1$, only relative phase information can be unveiled. In this context, it is useful to understand under which phase transformation $\phi_j\rightarrow\phi'_j$ the correlation function stays invariant. 

As only time and position differences between particle events are taken into account, the second-order correlation function is invariant under position shifts $y\rightarrow y+ y_0$ and time transformations $t\rightarrow t+ t_0$. Using equation (\ref{eq1}) and (\ref{eq2}) such a time transformation and position shift equals transformations
\begin{align}
y\rightarrow y+ y_0 \quad &\widehat{=} \quad \varphi(t)\rightarrow \varphi(t)+k y_0 ~,\label{eq48}\\
t\rightarrow t+ t_0 \quad &\widehat{=} \quad \phi_j\rightarrow \phi_j+\omega_j t_0 ~.\label{eq49}
\end{align}
Furthermore, the correlation function is invariant under simultaneous time and space reversal: $t\rightarrow -t\, ,\,y\rightarrow -y$, because only time and space differences are taken into account. According to equation (\ref{eq1}) and (\ref{eq2}), this transformation equals a phase transformation
\begin{equation}\label{eq50}
t\rightarrow -t\, ,\,y\rightarrow -y \quad \widehat{=} \quad \phi_j\rightarrow -\phi_j+\pi ~.
\end{equation}
Hence, under above phase transformations (equation (\ref{eq49}) and (\ref{eq50})) the correlation function stays unchanged. 

Exemplary, the invariance is calculated for a perturbation with two frequencies ($N=2$) and phases $\phi_1$ and $\phi_2$. According to equation (\ref{eq49}), the correlation function is invariant for following phase transformations
\begin{align}\label{eq51}
\hat{\phi}_1= \phi_1+\omega_1 t_0 \quad , \quad \hat{\phi}_2= \phi_2+\omega_2 t_0~.
\end{align}
Eliminating $t_0$ yields
\begin{equation}\label{eq52}
\hat{\phi}_2= \frac{\omega_2}{\omega_1}\hat{\phi}_1+\left(\phi_2 -\frac{\omega_2}{\omega_1}\phi_1\right) ~.
\end{equation}
Every phase duplet $\{\hat{\phi}_1,\hat{\phi}_2\}$ described by this equation, yields the same correlation function as the original phase duplet $\{\phi_1,\phi_2\}$. Therefore, the absolute phases $\phi_1$ and $\phi_2$ can not be uniquely determined by the correlation analysis. A second set of phase duplets $\{\tilde{\phi}_1,\tilde{\phi}_2\}$ with identical correlation functions can be found after evaluating the phase transformation for time and space reversal from equation (\ref{eq50})
\begin{equation}\label{eq53}
\tilde{\phi}_1= -\hat{\phi}_1+\pi  \quad , \quad \tilde{\phi}_2 = -\hat{\phi}_2+\pi~,
\end{equation}
yielding together with equation (\ref{eq52})
\begin{equation}\label{eq54}
\tilde{\phi}_2= \frac{\omega_2}{\omega_1}\tilde{\phi}_1+\left(\frac{\omega_2}{\omega_1}\phi_1-\phi_2 +\pi\left(1-\frac{\omega_2}{\omega_1}\right)\right) ~.
\end{equation} 
As result, there are two linear equations (equation (\ref{eq52}) and (\ref{eq54})), that describe the same correlation function. It is thus not possible to determine the original phases $\phi_1$ and $\phi_2$ from the correlation analysis. 

\subsubsection{Determination of perturbation phases}\label{sec2.8.2}
Although the original phases $\phi_j$ can not be determined by the correlation analysis, the identification is possible by the reconstruction of the unperturbed interference pattern using the obtained perturbation parameters $\omega_j$ and $\varphi_j$ \cite{Rembold2016}
\begin{equation}\label{eq55}
y_{i,new} = y_i - \frac{\lambda}{2\pi}\varphi(t_i) = y_i - \frac{\lambda}{2\pi}\sum_{j=1}^N \varphi_j \cos\left(\omega_j t_i + \phi_j\right)~, 
\end{equation} 
with the spatial coordinate of the reconstructed interference pattern $y_{i,new}$, the time-dependent perturbation $\varphi(t_i)$ (equation (\ref{eq2})) and $y_i$, $t_i$ the spatial and temporal coordinate of particle $i$ forming the perturbed interference pattern. The phases $\phi_j$ are varied until the maximum contrast of the reconstructed interference pattern is achieved. Using equation (\ref{eq52}) or (\ref{eq54}) in the reconstruction process, the free parameter space for the phases $\phi_j$ can be reduced by one dimension.

\section{Numerical second-order correlation function}\label{sec3}
In this chapter, the characteristics of the numerical second-order correlation function used for the evaluation of the experimental data are analyzed in detail. First, the numerical correlation function is derived and investigated how the discretization affects the obtained parameters as contrast and perturbation amplitude. A theoretical description of the influence on the correlation analysis and the effect of noise is presented. This theory is numerically cross-checked by simulations of single-particle interference. Afterwards, the possibility to use the correlation analysis for broad-band frequency noise is demonstrated by a simulation with a Gaussian distributed noise. At the end of this chapter, the application limits of the second-order correlation theory are discussed.

\subsection{Correlation function for discrete signals}\label{sec3.1}
So far, the correlation function has been calculated from the analytic probability distribution of particle impacts at the detector. In reality, however, the detector yields a discrete signal of particle impacts for a finite acquisition time $T$. The corresponding second-order correlation function $g^{(2)}(u,\tau)$ is thus extracted from this detector signal $f(y,t)$ consisting of $N$ particle arrival times $t_i$ and coordinates $y_i$, ($i=1\ldots N$)
\begin{equation}\label{eq56}
f(y,t) = \sum_{i=1}^N \delta\left(y-y_i\right)\delta\left(t-t_i\right)~.
\end{equation}
Here, $\delta$ denotes the Dirac delta function, which assures normalization of $f(y,t)$
\begin{equation}\label{eq57}
\int_{-\infty}^{\infty}\int_{-\infty}^{\infty} f(y,t)\,\mathrm{d}y\mathrm{d}t = N~.
\end{equation}
The approximation of the particle impacts in the detection plane by the Dirac delta function is suitable as the pulse width at the detector is much smaller than the mean time distance between particle impacts. Using this definition of the detector signal, the correlation function is calculated according to equation (\ref{eq9}). As before, the denominator is calculated first
\begin{align}\label{eq58}
\llangle f(y+u,t+\tau) \rrangle_{y,t}\,&=\,\frac{1}{TY}\int_{0}^{T}\int_{-Y/2}^{Y/2}\sum_{j=1}^N \delta\left(y+u-y_j\right)\delta\left(t+\tau-t_j\right)\,\mathrm{d}y\mathrm{d}t \nonumber\\ 
&=\,\frac{1}{TY}\sum_{j=1}^N\,\underbrace{\int_{0}^{T}\delta\left(t-t_j+\tau\right)\,\mathrm{d}t}_{=\,1}\underbrace{\int_{-Y/2}^{Y/2} \delta\left(y-y_j+u\right)\,\mathrm{d}y}_{=\,1} \nonumber\\ 
&=\,\frac{1}{TY}\sum_{j=1}^N 1 = \,\frac{N}{TY}~,
\end{align}
with the acquisition time $T$ and length $Y$. The second term in the denominator can be determined from equation (\ref{eq58}) for $u=0$ and $\tau=0$, yielding $\llangle f(y,t) \rrangle_{y,t}\,=\,N/TY$. Using equation (\ref{eq56}), the numerator in equation (\ref{eq9}) becomes
\begin{align}\label{eq59}
\llangle f(y+u,&t+\tau)f(y,t) \rrangle_{y,t}\,= \nonumber\\ 
&=\,\frac{1}{TY}\int_{0}^{T}\int_{-Y/2}^{Y/2}\sum_{i,j=1}^N \delta\left(y+u-y_j\right)\delta\left(t+\tau-t_j\right)\delta\left(y-y_i\right)\delta\left(t-t_i\right)\,\mathrm{d}y\mathrm{d}t \nonumber\\ 
&=\,\frac{1}{TY}\sum_{i,j=1}^N\,\underbrace{\int_{0}^{T}\delta\left(t+\tau-t_j\right)\delta\left(t-t_i\right)\,\mathrm{d}t}_{=\,\delta\left(\tau+t_i-t_j\right)}\underbrace{\int_{-Y/2}^{Y/2}\delta\left(y+u-y_j\right)\delta\left(y-y_i\right)\,\mathrm{d}y}_{=\,\delta\left(u+y_i-y_j\right)} \nonumber\\  
&=\,\frac{1}{TY}\sum_{i,j=1}^N \delta\left(\tau+t_i-t_j\right)\delta\left(u+y_i-y_j\right) ~.
\end{align}
With the results of equation (\ref{eq58}) and (\ref{eq59}), the discrete second-order correlation function is given by
\begin{align}\label{eq60}
g^{(2)}(u,\tau) &= \frac{\llangle f(y+u,t+\tau) f(y,t)\rrangle_{y,t}}{\llangle f(y+u,t+\tau) \rrangle_{y,t} \llangle f(y,t) \rrangle_{y,t}} \nonumber\\  &=\frac{TY}{N^2}\sum_{i,j=1}^N\delta\left(\tau+t_i-t_j\right)\delta\left(u+y_i-y_j\right)~.
\end{align} 
To achieve a proper correlation function, which can be compared with the analytic theory, it is necessary to implement a temporal and spatial discretization step size $\Delta\tau$ and $\Delta u$ in equation (\ref{eq60}), yielding
\begin{align}\label{eq61}
g^{(2)}(u,\tau) &= \frac{TY}{N^2}\frac{1}{\Delta\tau\Delta u}\underbrace{\int_{\tau-\frac{\Delta\tau}{2}}^{\tau+\frac{\Delta\tau}{2}}\int_{u-\frac{\Delta u}{2}}^{u+\frac{\Delta u}{2}}\sum_{i,j=1}^N \delta\left(\tau+t_i-t_j\right)\delta\left(u+y_i-y_j\right)\,\mathrm{d}u\mathrm{d}\tau}_{=\,N_{u,\tau}} \nonumber\\ 
& =\frac{TY}{N^2\Delta\tau\Delta u} N_{u,\tau}~,
\end{align}
with $N_{u,\tau}$ denoting the number of particle pairs $(i,j)$ with distances $(y_i-y_j)\in[u-\Delta u/2,u+\Delta u/2]$ and time separations $(t_i-t_j)\in[\tau-\Delta \tau/2,\tau+\Delta \tau/2]$.

Due to the finite acquisition time $T$ and length $Y$, the probability to detect particle impacts with large temporal and spatial differences is reduced by a factor of $\left(1-\tau/T\right)$ and $\left(1-|u|/Y\right)$, respectively. Therefore, the number $N_{u,\tau}$ in equation (\ref{eq61}) has to be corrected for this effect, resulting in
\begin{equation}\label{eq62}
g^{(2)}(u,\tau) = \frac{TY}{N^2\Delta\tau\Delta u}\frac{N_{u,\tau}}{\Big(1-\frac{\tau}{T}\Big)\Big(1-\frac{|u|}{Y}\Big)} ~.
\end{equation}
As with the analytic correlation function, the numerical correlation function in equation (\ref{eq62}) is normalized to one.
\begin{figure}
\centering
\includegraphics[width=1\textwidth]{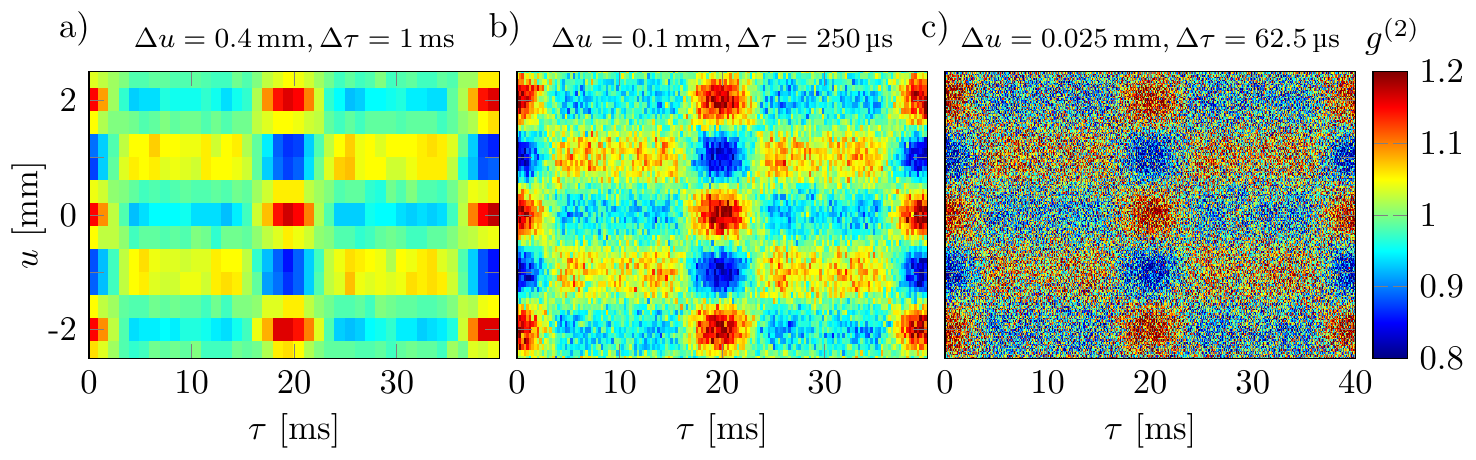} \caption{Correlation functions with different temporal and spatial discretization step sizes $\Delta u$ and $\Delta\tau$ extracted according to equation (\ref{eq62}) from a single-particle simulation of an interference pattern with a contrast $K=0.6$ and a spatial periodicity $\lambda=\unit[2]{mm}$ perturbed with $\omega_1/2\pi=\unit[50]{Hz}$ and $\varphi_1=\unit[0.75]{\pi}$. a) The structure is ``smeared out", because the discretization step size is too large. b) Compared to a), the structure is clearly visible, but the noise increases with smaller $\Delta u$ and $\Delta\tau$. c) For even smaller discretization step size, the noise grows larger.}
\label{fig6}
\end{figure}

Three correlation functions with different spatial and temporal discretization step sizes are illustrated in figure \ref{fig6}. They are extracted using equation (\ref{eq62}) from a single-particle simulation of an interference pattern (section \ref{sec3.3}) with a contrast $K=0.6$ and a spatial periodicity $\lambda=\unit[2]{mm}$ perturbed with $\omega_1/2\pi=\unit[50]{Hz}$ and $\varphi_1=\unit[0.75]{\pi}$. If $\Delta u$ and $\Delta\tau$ are too large, the structure in the correlation function gets ``smeared out" as can be seen in figure \ref{fig6}(a). In contrast, the structure is clearly visible in figure \ref{fig6}(b). However, the noise in the correlation function increases compared to figure \ref{fig6}(a). This noise grows larger for even smaller discretization step size as seen in figure \ref{fig6}(c). Therefore, it is necessary to understand the effects of discretization and choose $\Delta u$ and $\Delta\tau$ in an appropriate way.

\newpage

\subsection{Discretization effects}\label{sec3.2}
In this section, the effects of the discretization step size on the correlation analysis and noise in the correlation function are analyzed. Therefore, the influence on the contrast and peak phase deviation due to the discretization of the correlation function is derived for one perturbation frequency. The theory reveals how the spatial and temporal discretization step size have to be chosen. Afterwards, a theoretical description for the noise in the correlation function and the corresponding amplitude spectrum is derived. Thereby, an optimum spatial discretization step size is found, resulting in a maximum signal-to-noise ratio. At the end, an estimation of the smallest detectable peak phase deviation is given.  

\subsubsection{Influence on correlation analysis}\label{sec3.2.1}
To study the influence of the spatial and temporal discretization step size $\Delta u$ and $\Delta\tau$, the analytic solution from equation (\ref{eq34}) is used. For simplicity, only one perturbation frequency $\omega_1$ is taken into account. Discretizing the correlation function into temporal and spatial intervals $\Delta\tau$ and $\Delta u$, then yields
\begin{align}\label{eq63}
g^{(2)}(u,&\tau)_{\Delta u,\Delta\tau}=\frac{1}{\Delta\tau\Delta u}\int_{u-\frac{\Delta u}{2}}^{u+\frac{\Delta u}{2}}\int_{\tau-\frac{\Delta\tau}{2}}^{\tau+\frac{\Delta\tau}{2}} g^{(2)}(u,\tau)\, \mathrm{d}\tau\mathrm{d}u \nonumber\\ 
&= 1 + \frac{K^2}{2}\frac{1}{\Delta u}\int_{u-\frac{\Delta u}{2}}^{u+\frac{\Delta u}{2}}\cos\left(ku\right)\mathrm{d}u\,\cdot\sum_{m_1=-\infty}^\infty J_{m_1}\left(\varphi_1\right)^2\frac{1}{\Delta\tau}\int_{\tau-\frac{\Delta\tau}{2}}^{\tau+\frac{\Delta\tau}{2}} \cos\left(m_1\omega_1\tau\right)\mathrm{d}\tau \nonumber\\ 
&=1 + \frac{K^2}{2}\cdot A\left(\tau\right)_{\Delta\tau}\sinc\left(k\frac{\Delta u}{2}\right)\cos\left(ku\right) ~,
\end{align}
with the $\sinc$-function $\sinc(x)=\sin(x)/x$ and the amplitude
\begin{equation}\label{eq64}
A\left(\tau\right)_{\Delta\tau} =\, \sum_{m_1=-\infty}^\infty J_{m_1}\left(\varphi_1\right)^2\cdot\sinc\left(m_1\omega_1\frac{\Delta\tau}{2}\right)\cos\left(m_1\omega_1\tau\right)~.
\end{equation}
For $\Delta u=0$ and $\Delta\tau=0$, equation (\ref{eq63}) and (\ref{eq64}) yield the analytic solution of the correlation function in equation (\ref{eq34}) and (\ref{eq35}) for $N=1$. 

\begin{figure}
\centering
\includegraphics[width=0.9\textwidth]{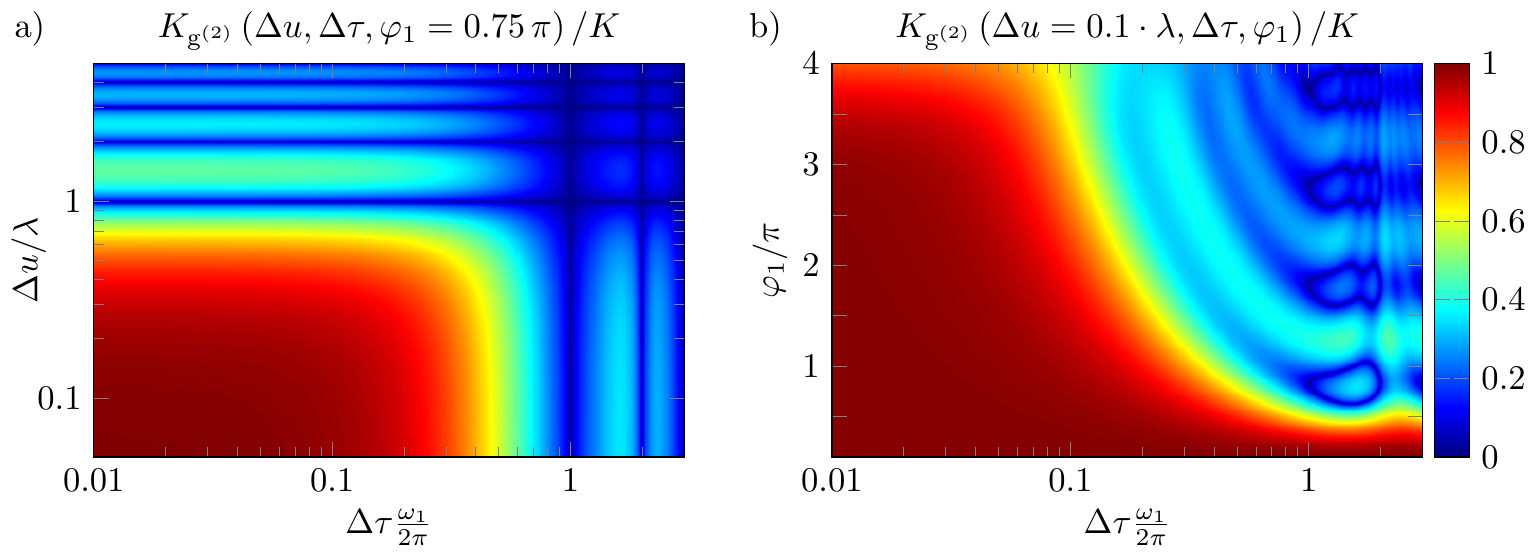} \caption{a) Dependence of the extracted contrast $K_{\text{g}^{(2)}}\left(\Delta u,\Delta\tau,\varphi_1=\unit[0.75]{\pi}\right)/K$ on the spatial and temporal discretization step size $\Delta u$ and $\Delta\tau$ for one perturbation frequency $\omega_1$ with a peak phase deviation of $\varphi_1=\unit[0.75]{\pi}$. b) Dependence of $K_{\text{g}^{(2)}}\left(\Delta u=0.1\cdot\lambda,\Delta\tau,\varphi_1\right)/K$ on the temporal discretization $\Delta\tau$ and peak phase deviation $\varphi_1$ for $\Delta u/\lambda = 0.1$.}
\label{fig7}
\end{figure}

As discussed in section \ref{sec2.4}, the contrast $K$ and spatial periodicity $\lambda$ of the unperturbed interference pattern are determined from the correlation function at $\tau=0$ (equation (\ref{eq36})). For this correlation time, equation (\ref{eq63}) and (\ref{eq64}) become
\begin{equation}\label{eq65}
g^{(2)}(u,0)_{\Delta u,\Delta\tau}=1 + \frac{K^2}{2}\cdot A\left(0\right)_{\Delta\tau}\sinc\left(k\frac{\Delta u}{2}\right)\cos\left(ku\right) ~,
\end{equation}
with
\begin{equation}\label{eq66}
A(0)_{\Delta\tau} = \sum_{m_1=-\infty}^\infty J_{m_1}\left(\varphi_1\right)^2\cdot\sinc\left(m_1\omega_1\frac{\Delta\tau}{2}\right)~.
\end{equation}
Thus, the spatial periodicity $\lambda=2\pi/k$ is not influenced by the discretization step size, but the extracted contrast $K_{\text{g}^{(2)}}$ is modified to 
\begin{equation}\label{eq67}
K_{\text{g}^{(2)}}\left(\Delta u,\Delta\tau,\varphi_1\right)=K\cdot\sqrt{\Big|A(0)_{\Delta\tau}\cdot\sinc\left(k\frac{\Delta u}{2}\right)\Big|}~.
\end{equation}
This contrast is zero for multiples of $\Delta\tau=2\pi/\omega_1$ and $\Delta u=\lambda$, because the $\sinc$-function is zero for this values. For $\Delta\tau\rightarrow 0$ and $\Delta u\rightarrow 0$, the $\sinc$-functions in equation (\ref{eq66}) and (\ref{eq67}) approach to 1. Full contrast $K$ is only achieved for $\Delta u=\Delta\tau=0$. The dependence of the extracted contrast $K_{\text{g}^{(2)}}$ on $\Delta\tau$ and $\Delta u$ is illustrated in figure \ref{fig7}(a) for a peak phase deviation of $\varphi_1=\unit[0.75]{\pi}$. The dependence of the extracted contrast on the temporal discretization step size $\Delta\tau$ and the peak phase deviation $\varphi_1$ is shown in figure \ref{fig7}(b) for $\Delta u/\lambda = 0.1$.

To determine the perturbation frequency $\omega_1$ and corresponding peak phase deviation $\varphi_1$ from the correlation function, the analytic amplitude spectrum of equation (\ref{eq47}) is used. For discrete signals, the amplitude spectrum also depends on $\Delta u$ and $\Delta\tau$. This dependence is calculated from equation (\ref{eq63}) and (\ref{eq64}) with a temporal Fourier transformation, yielding
\begin{align}\label{eq68}
\frac{1}{2\pi}\Big|\mathcal{F}&\left(g^{(2)}(u,\tau)_{\Delta u,\Delta\tau}\right)(u,\omega)\Big|^2 = \delta(\omega)^2+\Bigg(\frac{K^2}{2}\sinc\left(k\frac{\Delta u}{2}\right)\cos\left(ku\right)\cdot\\ \nonumber
&\cdot\sum_{m_1=-\infty}^\infty J_{m_1}(\varphi_1)^2\cdot\sinc\left(m_1\omega_1\frac{\Delta\tau}{2}\right)\bigg(\delta(\omega+m_1\omega_1)+\delta(\omega-m_1\omega_1)\bigg)\Bigg)^2 ~, \nonumber
\end{align}
with the frequency components $\pm m_1\omega_1$. Comparing equation (\ref{eq68}) with the analytic solution in equation (\ref{eq47}) for $N=1$, it can be seen, that the frequency components $\pm m_1\omega_1$ are not changed, but their amplitudes $J_{m_1}(\varphi_1)^2$ are modified in equation (\ref{eq68}) due to the spatial and temporal discretization. This dependence is calculated for the amplitude of the fundamental frequency component in equation (\ref{eq68}) with $m_1=1$ and $u=0$, resulting in
\begin{equation}\label{eq69}
\frac{1}{\sqrt{2\pi}}\Big|\mathcal{F}\left(g^{(2)}(0,\tau)_{\Delta u,\Delta\tau}\right)(0,\omega_{\{m_1=1\}})\Big| = \frac{K^2}{2}\,J_{1}(\bar{\varphi}_1)^2~,
\end{equation}
with the modified amplitude $J_{1}(\bar{\varphi}_1)^2$ of the fundamental frequency component $\omega_1$ depending on $\Delta u$ and $\Delta\tau$
\begin{equation}\label{eq70}
J_{1}(\bar{\varphi}_1)^2 = \sinc\left(k\frac{\Delta u}{2}\right)\sinc\left(\omega_1\frac{\Delta\tau}{2}\right)\cdot J_{1}(\varphi_1)^2~.
\end{equation}
Here, the amplitude of the fundamental frequency component $J_{1}(\varphi_1)^2$ is reduced by the factor of $\sinc\left(k\frac{\Delta u}{2}\right)\sinc\left(\omega_1\frac{\Delta\tau}{2}\right)$, which needs to be taken into account for the determination of $\varphi_1$. Therefore, the peak phase deviation $\bar{\varphi}_1$ extracted from the amplitude spectrum of the correlation function also depends on the discretization step size via the square of the Bessel function. For $\Delta u\rightarrow 0$ and $\Delta\tau\rightarrow 0$, equation (\ref{eq69}) results in $K^2/2\cdot J_{1}(\varphi_1)^2$ yielding the correct peak phase deviation $\varphi_1$.

\subsubsection{Influence of noise}\label{sec3.2.2}
As discussed in section \ref{sec3.1}, a real experiment requires the correlation function to be derived from a finite number of detection events. More precisely, it is calculated from the number of correlated particle pairs $N_{u,\tau}$ within a given correlation window $\Delta\tau$ and $\Delta u$. Due to the statistical nature of the particles, $N_{u,\tau}$ is subject to Poissonian noise, which is transferred onto the correlation function and the corresponding amplitude spectrum. In principle, this limits the signal-to-noise ratio and the minimal detectable perturbation amplitude.
 
In the following section, this effect is estimated and optimal settings for the discretization step size are found. Therefore, all fluctuating variables $X$ are described by their corresponding mean values $\big\langle X \big\rangle$ and variances $\sigma^2_X$, with $\sigma_X$ being the standard deviation
\begin{equation}\label{eq71}
\sigma^2_X=\Big\langle \big(X - \langle X \rangle\big)^2\Big\rangle = \big\langle X^2\big\rangle - \big\langle X\big\rangle^2~.
\end{equation}
For simplicity reasons the analysis is restricted to correlation times $\tau\ll T$ and positions $u\ll Y$. With the correlation function from equation (\ref{eq62}) being normalized to 1, it can be found
\begin{equation}\label{eq72}
g^{(2)}(u,\tau)=\frac{N_{u,\tau}}{\big\langle N_{u,\tau}\big\rangle} \quad \text{with}\quad \big\langle N_{u,\tau} \big\rangle = \frac{N^2 \Delta\tau \Delta u}{TY}~.
\end{equation}
The expected standard deviation can then be calculated yielding
\begin{equation}\label{eq73}
\sigma^2_{g^{(2)}} = \Big\langle g^{(2)}(u,\tau)^2 \Big\rangle - \Big\langle g^{(2)}(u,\tau) \Big\rangle^2 = \frac{\big\langle N_{u,\tau}^2\big\rangle}{\big\langle N_{u,\tau}\big\rangle^2}- 1~.
\end{equation}
With $N_{u,\tau}$ following a Poissonian distribution, variance and mean value are directly linked
\begin{equation}\label{eq74}
\sigma^2_{N_{u,\tau}}=\big\langle N_{u,\tau}^2 \big\rangle - \big\langle N_{u,\tau} \big\rangle^2 = \big\langle N_{u,\tau}\big\rangle~,
\end{equation}
yielding
\begin{equation}\label{eq75}
\sigma^2_{g^{(2)}} = \frac{1}{\big\langle N_{u,\tau}\big\rangle}=\frac{TY}{N^2\Delta\tau \Delta u} = \frac{N_\tau N_u}{N^2}~.
\end{equation}
As expected, the variance (noise) of the correlation function depends on the total number of detected particles and the number of bins $N_\tau=T/\Delta\tau$ and $N_u=Y/\Delta u$ in the temporal and spatial direction.

Before calculating, how the noise in the correlation function transfers onto its amplitude spectrum, the correlation function is split in two parts: The first part $g^{(2)}_{id}$ describing the ideal correlation function, as expected in the limit of infinite detection events and the second part $f$ describing the noise only
\begin{equation}\label{eq76}
g^{(2)}(\tau) = g^{(2)}_{id}(\tau) + f(\tau)~.
\end{equation}
Obviously, mean values and standard deviations of these functions are given to
\begin{equation}\label{eq77}
\Big\langle g^{(2)}_{id}\Big\rangle=\Big\langle g^{(2)}\Big\rangle,\, \sigma_{g^{(2)}_{id}}=0  \quad\text{and}\quad\big\langle f(\tau)\big\rangle = 0,\, \sigma_f = \sigma_{g^{(2)}}~.
\end{equation}
Following this description, the amplitude spectrum of the correlation function reads
\begin{equation}\label{eq78}
\big|\mathcal{G}(\omega)\big| = \big|\mathcal{G}_{id}(\omega)+\mathcal{F}(\omega)\big| \approx \big|\mathcal{G}_{id}(\omega)\big| + \big|\mathcal{F}(\omega)\big|~,
\end{equation}
with $\mathcal{G}$ and $\mathcal{F}$ denoting the discrete Fourier transforms of the time discrete signals $g^{(2)}$ and $f$
\begin{equation}\label{eq79}
\mathcal{F}(\omega)=\frac{1}{\tau_{\text{max}}}\int_0^{\tau_{\text{max}}} f(\tau) e^{i\omega \tau}d\tau = \frac{\Delta\tau}{\tau_{\text{max}}}\sum_{i=1}^N f_n e^{i\omega t_n}~.
\end{equation}
Here, $\tau_{\text{max}}$ denotes the maximum correlation time up to which the correlation function is evaluated. The noise in the spectrum is thus solely included in $\left|\mathcal{F}(\omega)\right|$. Using Parsival's theorem \cite{arfken1999} together with $N_{\tau_{\text{max}}}=\tau_{\text{max}}/\Delta\tau$, it can be found
\begin{equation}\label{eq80}
\Big\langle\big| \mathcal{F}(\omega)\big|^2\Big\rangle = \frac{\sigma^2_f}{N_{\tau_{\text{max}}}} \quad \text{and} \quad \Big\langle\big| \mathcal{F}(\omega)\big|\Big\rangle^2 = \frac{\pi}{4}\frac{\sigma^2_f}{N_{\tau_{\text{max}}}}
\end{equation}
and thus a direct link between the noise in the power spectrum and the noise in the correlation function (equation (\ref{eq75}))
\begin{equation}\label{eq81}
\sigma^2_{\left|\mathcal{F}(\omega)\right|} = \left(1-\frac{\pi}{4}\right)\frac{\sigma^2_{g^{(2)}}}{N_{\tau_{\text{max}}}} = \left(1-\frac{\pi}{4}\right)\frac{N_u}{N^2}\frac{T}{\tau_{\text{max}}}~.
\end{equation} 

\begin{figure}
\centering
\includegraphics[width=0.5\textwidth]{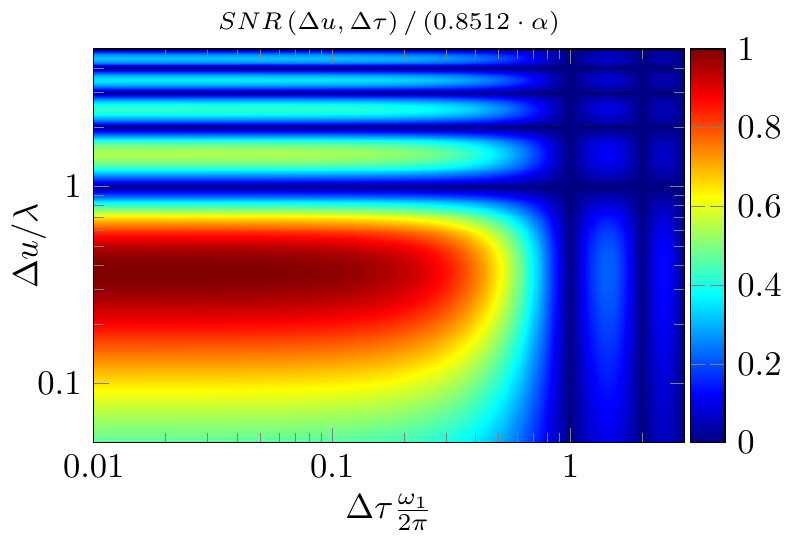} \caption{Signal-to-noise ratio calculated with equation (\ref{eq82}) and normalized to $\alpha$ (equation (\ref{eq83})) and $f\left(k\frac{\Delta u_{\text{opt}}}{2}\right)=0.8512$. The optimum spatial discretization can be identified at $\Delta u/\lambda = 0.371$.}
\label{fig8}
\end{figure}

The signal-to-noise ratio $SNR\left(\Delta u,\Delta\tau \right)$ of the amplitude spectrum used for the determination of the peak phase deviation $\varphi_1$ can be calculated with equation (\ref{eq69}), (\ref{eq70}) and (\ref{eq81}) yielding
\begin{align}\label{eq82}
SNR\left(\Delta u,\Delta\tau \right) &= \frac{\frac{1}{\sqrt{2\pi}}\Big|\mathcal{F}\left(g^{(2)}(0,\tau)_{\Delta u,\Delta\tau}\right)(0,\omega_{\{m_1=1\}})\Big|}{\sigma_{\left|\mathcal{F}(\omega)\right|}} \nonumber\\ 
&= \frac{K^2 J_{1}(\varphi_1)^2}{\sqrt{\left(1-\frac{\pi}{4}\right)2k}}\frac{\sin\left(k\frac{\Delta u}{2}\right)}{\left(k\frac{\Delta u}{2}\right)^{\frac{1}{2}}} \sinc\left(\omega_1\frac{\Delta\tau}{2}\right)\cdot \frac{N\sqrt{\tau_{\text{max}}/T}}{\sqrt{Y}} \nonumber\\ 
&= \alpha\cdot\frac{\sin\left(k\frac{\Delta u}{2}\right)}{\left(k\frac{\Delta u}{2}\right)^\frac{1}{2}} \sinc\left(\omega_1\frac{\Delta\tau}{2}\right)~,  
\end{align}
with
\begin{equation}\label{eq83}
\alpha = 0.6089\cdot K^2 J_{1}(\varphi_1)^2\cdot \frac{N\sqrt{\tau_{\text{max}}/T}}{\sqrt{Y/\lambda}}~.
\end{equation}
For $\Delta\tau\rightarrow 0$, the $\sinc$-function approaches to 1 and the signal-to-noise ratio only depends on the spatial discretization step size $\Delta u$. The function $f\left(k\frac{\Delta u}{2}\right)=\sin\left(k\frac{\Delta u}{2}\right)/\left(k\frac{\Delta u}{2}\right)^{\frac{1}{2}}$ has a global maximum at the position $k\frac{\Delta u_{\text{opt}}}{2}=1.1656$ with the maximum value of $f\left(k\frac{\Delta u_{\text{opt}}}{2}\right)=0.8512$. Hence, the optimum spatial discretization step size, leading to a maximum of the signal-to-noise ratio, becomes
\begin{equation}\label{eq84}
\Delta u_{\text{opt}}=0.371\cdot\lambda~.
\end{equation}
The signal-to-noise ratio calculated with equation (\ref{eq82}) is plotted in figure \ref{fig8} for different spatial and temporal discretization step sizes. The optimum spatial discretization step size is clearly visible at $\Delta u/\lambda = 0.371$. From equation (\ref{eq82}) and (\ref{eq83}), the optimum signal-to-noise ratio is deduced for $\Delta u_{\text{opt}}=0.371\cdot\lambda$ and $\Delta\tau\rightarrow 0$
\begin{equation}\label{eq85}
SNR_{\text{opt}}\left(\Delta u_{\text{opt}}=0.371\cdot\lambda,\Delta\tau\rightarrow 0 \right) = 0.5183\cdot K^2 J_{1}(\varphi_1)^2\cdot\frac{N\sqrt{\tau_{\text{max}}/T}}{\sqrt{Y/\lambda}}~.
\end{equation}
A lower limit for the identification of small peak phase deviations $\varphi_1\ll 1$ can be derived from equation (\ref{eq85}) by setting $SNR_{\text{opt}}$ equal to 1. This sets the threshold at which the noise and signal have equal amplitude. With $J_{1}(\varphi_1)^2\approx\varphi_1^2/4$ and equation (\ref{eq85}), this yields
\begin{equation}\label{eq86}
\varphi_{1,\text{min}}=0.8842\,\pi\cdot\frac{\left(Y/\lambda\right)^{\frac{1}{4}}}{KN^{\frac{1}{2}}\left(\tau_{\text{max}}/T\right)^{\frac{1}{4}}}~.
\end{equation}
For the measurement of small peak phase deviations it is thus favourable to have an interference pattern with large contrast $K$ and pattern periodicity $\lambda$. Furthermore, a large number of particles $N$ decreases the minimum detectable peak phase deviation.

\subsection{Simulations of single-particle interference}\label{sec3.3}
To numerically cross-check the theoretical calculations in section \ref{sec3.2}, a set of $N$ particles with temporal and spatial coordinates $(t_i,y_i),i=1\ldots N$ have been simulated. Therefore, time and position coordinates are generated according to the corresponding distribution functions using the acceptance-rejection method \cite{casella2004generalized}. For the temporal coordinates, the probability distribution of time differences $\Delta t_i =t_{i+1}-t_i$ between successive events is used, which for Poisson statistics is given by \cite{haight1967handbook}
\begin{equation}\label{eq87}
p(\Delta t)=\mbox{e}^{-cr\cdot \Delta t}~.
\end{equation}
Here, $cr=N/T$ denotes the mean count rate. Following this distribution function, a set of $N-1$ time differences $\Delta t_i, i=1\ldots N-1$ is generated. Starting with the first event at $t_1=0$, the time steps for successive events are given by
\begin{equation}\label{eq88}
t_j=\sum_{i=1}^{j-1} \Delta t_i~.
\end{equation}
After having generated all time coordinates $t_i$, the corresponding spatial coordinates $y_i$ are created according to the probability distribution
\begin{equation}\label{eq89}
p(y,t_i)=1+K\cos \big(ky + \varphi\left(t_i\right)\big)~,
\end{equation}
with the time-dependent perturbation $\varphi\left(t_i\right)$ from equation (\ref{eq2}). This results in a full set of time and position coordinates.
\begin{figure}
\centering
\includegraphics[width=1\textwidth]{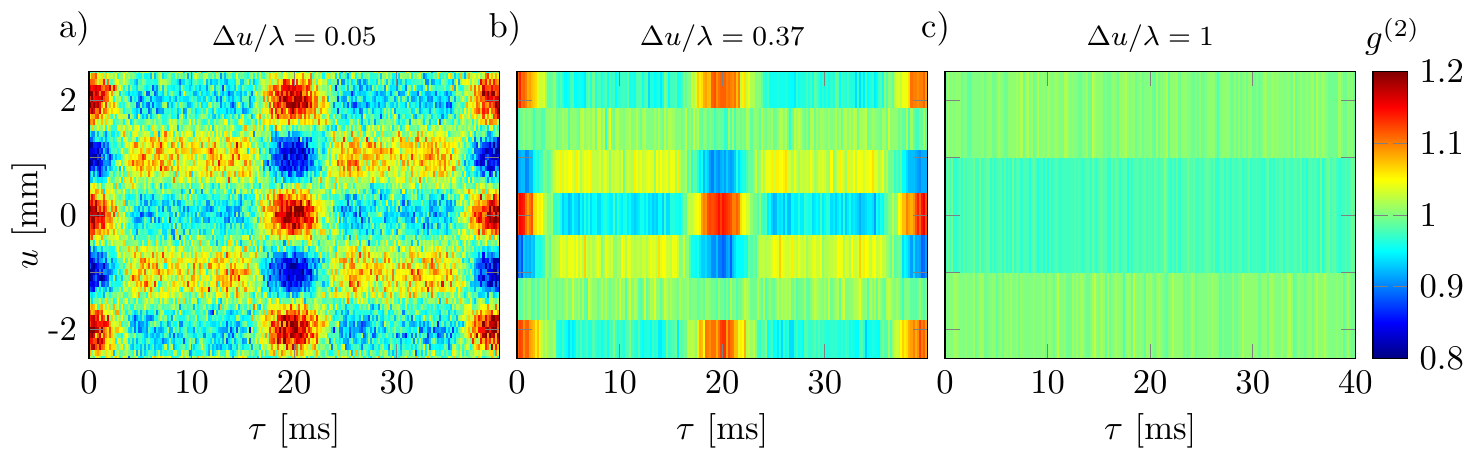} \caption{Three correlation functions for different spatial $\Delta u/\lambda$ and a fixed temporal discretization step size $\Delta\tau\omega_1/2\pi=0.01$. They are extracted from a single-particle simulation of an interference pattern with $K=0.6$ and $\lambda=\unit[2]{mm}$ perturbed with $\omega_1/2\pi=\unit[50]{Hz}$ and $\varphi_1=\unit[0.4]{\pi}$ used in equation (\ref{eq89}). The spatial discretization step size is increased from a) to c) and therefore the structure of the correlation function is ``smeared out" until it is vanished in c). Otherwise, the noise decreases for larger $\Delta u/\lambda$.}
\label{fig9}
\end{figure}

Such a simulation has been made for an interference pattern with a contrast of $K=0.6$, a spatial periodicity of $\lambda=\unit[2]{mm}$ and a single frequency perturbation $\varphi\left(t\right)$ with $\omega_1/2\pi=\unit[50]{Hz}$ and $\varphi_1=\unit[0.4]{\pi}$. With an acquisition time of $T=\unit[39.05]{s}$ and length $Y=\unit[20]{mm}$, $1.95\cdot 10^5$ particles have been simulated. This parameters have been chosen in accordance to typical experimental parameters \cite{Rembold2016}. The correlation function was extracted from the simulated data according to equation (\ref{eq62}) for different spatial discretization step sizes ($\Delta u$), a temporal discretization step size ($\Delta\tau=\unit[0.2]{ms}$) and a maximum correlation time of $\tau_{\text{max}}=\unit[1]{s}$. The temporal discretization step size is chosen to ensure that it does not reduce the signal-to-noise ratio (c.f. figure \ref{fig8} for $\Delta\tau\frac{\omega_1}{2\pi}=0.01$). Three correlation functions for different spatial discretization step sizes $\Delta u$ are illustrated in figure \ref{fig9}. With increasing $\Delta u$ from figure \ref{fig9}(a) to (c), the structure in the correlation function is ``smeared out". Otherwise, the noise decreases for a larger spatial discretization step size, because the mean particle number increases (equation (\ref{eq75})). The relation between this two effects leads to an optimum $\Delta u_{opt}$ (equation (\ref{eq84})) that provides a maximum of the signal-to-noise ratio (see equation (\ref{eq85})). 

\begin{figure}
\centering
\includegraphics[width=1\textwidth]{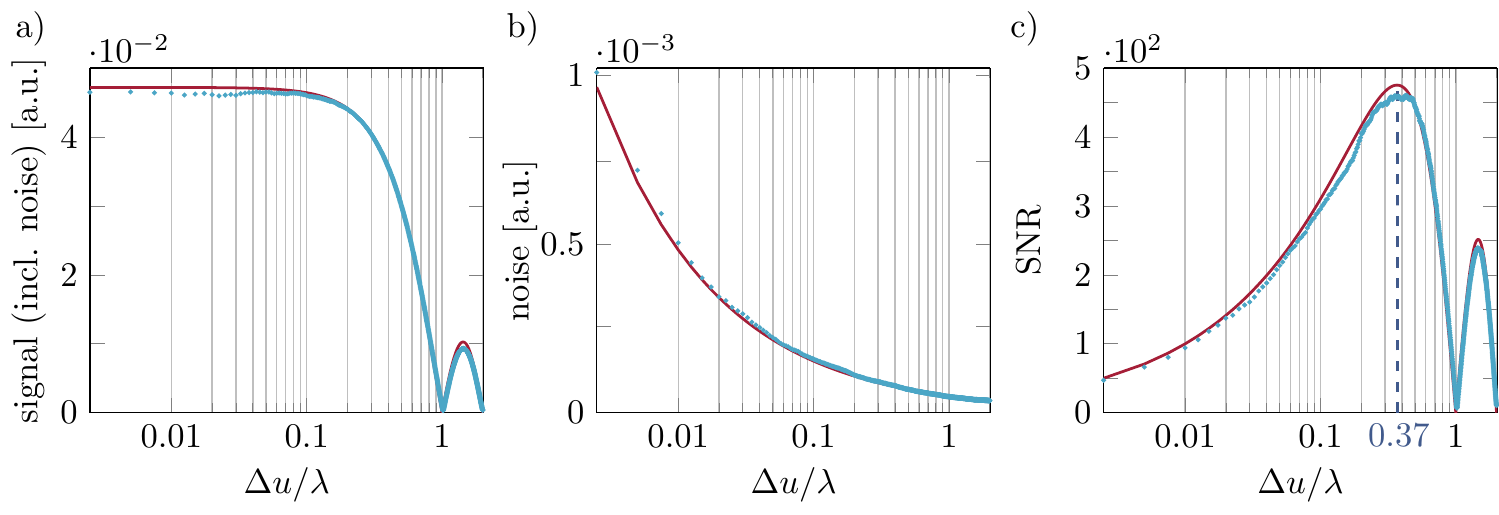} \caption{a) Signal height of the peak at the $\unit[50]{Hz}$ position in the amplitude spectrum of the correlation function (blue dots). The solid red line indicates the theoretical signal height calculated with equation (\ref{eq69}). b) The standard deviation of the noise determined from the amplitude spectrum. The theoretical curve was evaluated using equation (\ref{eq81}). c) The signal-to-noise ratio of the data points in a) and b) shows a good agreement to the theory calculated according to equation (\ref{eq82}) and (\ref{eq83}). The optimum spatial discretization can be identified at $\Delta u_{\text{opt}}/\lambda=0.37$.}
\label{fig10}
\end{figure}

As the correlation functions in figure \ref{fig9}, many correlation functions with different spatial discretisation step sizes ranging from $\Delta u =0.009\lambda$ to $\Delta u =2\lambda$ are extracted from the above simulation according to equation (\ref{eq62}). For each correlation function, the amplitude spectrum was calculated at the spatial position $u=\unit[0]{mm}$ using a numerical Fourier transformation. The signal height of the peak at the $\unit[50]{Hz}$ position and the standard deviation of the noise were determined in each spectrum. The result for the extracted signal height is plotted in figure \ref{fig10}(a) (blue dots). The standard deviation of the noise is shown in \ref{fig10}(b) and the signal-to-noise ratio in figure \ref{fig10}(c). The theoretical curve in figure \ref{fig10}(a) (red solid line) was calculated using equation (\ref{eq69}) with the parameters of the simulation. For $\Delta u/\lambda > 0.1$, the signal height is significantly reduced, because the structure in the correlation functions begins to ``smear out" (see figure \ref{fig9}(b)) until it is totally vanished for $\Delta u/\lambda=1$ (see figure \ref{fig9}(c)). The theoretical curve in figure \ref{fig10}(b) is evaluated with equation (\ref{eq81}) indicating that the noise is reduced for larger $\Delta u/\lambda$ due to the increasing mean particle number $\left< N_{u,\tau} \right>$ in equation (\ref{eq75}). The theoretical signal-to-noise ratio in figure \ref{fig10}(c) is calculated according to equation (\ref{eq82}) and (\ref{eq83}) illustrating the predicted optimum at $\Delta u_{\text{opt}}/\lambda=0.37$. The corresponding correlation function is shown in figure \ref{fig9}(b). The minimum peak phase deviation in equation (\ref{eq86}) can be calculated with the above values and yields $\varphi_{1,\text{min}}=\unit[1.49\cdot 10^{-2}]{\pi}$.

\subsection{Correlation analysis of broad-band frequency noise}\label{sec3.4}
To demonstrate the possibility, using the correlation analysis to describe not only single perturbation frequencies but also broad-band noise spectra, the following single-particle simulation has been made. The temporal coordinates $t_i$ of the particles are generated in the same way as described in section \ref{sec3.3}. The perturbation caused by a broad-band frequency noise is given by the corresponding amplitude spectrum $\hat{\varphi}\left(\omega_j\right)$ and phase spectrum $\hat{\Phi}\left(\omega_j\right)$. This is different to the former simulation, where the time-dependent perturbation $\varphi\left(t_i\right)$ was given by equation (\ref{eq2}). With the amplitude and phase spectrum, the time-dependent perturbation $\varphi\left(t_i\right)$ for the temporal coordinate $t_i$ can be calculated using a discrete Fourier transformation
\begin{equation}\label{eq90}
\varphi\left(t_i\right)= \frac{1}{\sqrt{2\pi}N_{\omega}}\sum_{j=1}^{N_{\omega}} \hat{\varphi}\left(\omega_j\right)\mbox{e}^{i\hat{\Phi}\left(\omega_j\right)}\mbox{e}^{-i\omega_j t_i}~,
\end{equation}
with the number of frequencies in the spectrum $N_{\omega}$. The spatial coordinate $y_i$ is determined as before according to the probability distribution (equation (\ref{eq89})) with the calculated phase shift of equation (\ref{eq90}). For the simulation demonstrated here, a Gaussian distributed noise spectrum with uncorrelated phases is applied. The discrete amplitude spectrum is thus given by
\begin{equation}\label{eq91}
\hat{\varphi}\left(\omega_j\right)=\varphi_0\cdot\mbox{e}^{-\frac{1}{2}\left(\frac{\omega_j -\omega_0}{\sigma_\omega}\right)^2}
\end{equation}
and the phase spectrum $\hat{\Phi}\left(\omega_j\right)$ is randomly distributed between $-\pi$ and $\pi$. Here, $\varphi_0$ denotes the maximum peak phase deviation, $\omega_0$ the central frequency and $\sigma_\omega$ the frequency standard deviation (band width). 
\begin{figure}
\centering
\includegraphics[width=0.5\textwidth]{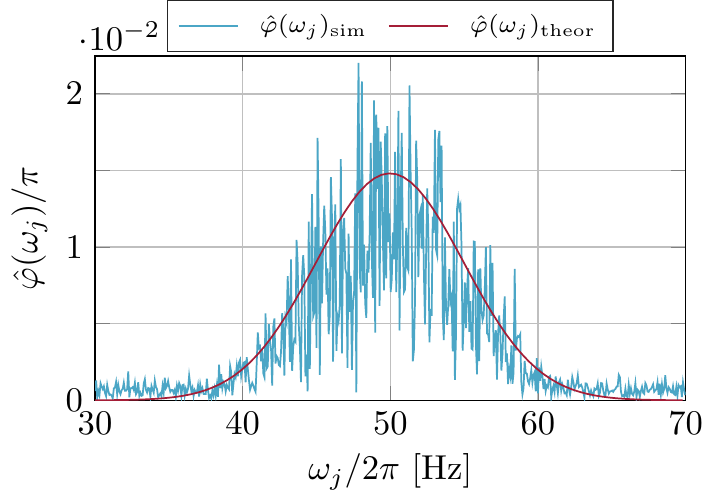} \caption{The blue solid line indicates the amplitude spectrum $\hat{\varphi}\left(\omega_j\right)_{\text{sim}}$. It is calculated with a numerical Fourier transformation from the time-dependent perturbation $\varphi\left(t_i\right)$ (equation (\ref{eq90})) used for the simulation in equation (\ref{eq89}). Theoretical amplitude spectrum $\hat{\varphi}\left(\omega_j\right)_{\text{theor}}$ (red solid line) resulting from the fit of equation (\ref{eq92}) to the correlation function extracted from the simulated data. The determined characteristics of the Gaussian distributed noise in equation (\ref{eq91}) are $\varphi_0=\unit[(1.48\pm 0.02)\cdot 10^{-2}]{\pi}$, $\omega_0/2\pi=\unit[(49.97\pm 0.11)]{Hz}$ and $\sigma_\omega/2\pi=\unit[(5.01\pm 0.11)]{Hz}$.}
\label{fig11}
\end{figure}

An interference pattern consisting of $5\cdot 10^5$ particles acquired in \unit[100]{s} with a contrast of $K=0.6$ and a spatial periodicity $\lambda=\unit[2]{mm}$ was perturbed by a Gaussian distributed noise according to equation (\ref{eq91}) with $\varphi_0=\unit[2\cdot 10^{-2}]{\pi}$, $\omega_0/2\pi=\unit[50]{Hz}$ and $\sigma_\omega/2\pi=\unit[5]{Hz}$. The frequency spectrum used for the simulation ranges from \unit[30]{Hz} to \unit[70]{Hz} with a resolution of \unit[1]{mHz}. After the simulation, the amplitude spectrum $\hat{\varphi}\left(\omega_j\right)_{\text{sim}}$ of the created time-dependent perturbation (equation (\ref{eq90})) is calculated using a numerical Fourier transformation and plotted in figure \ref{fig11} (blue solid line). The noise on the Gaussian distribution originates from the finite acquisition time and the randomly chosen phases.

\begin{figure}
\centering
\includegraphics[width=0.7\textwidth]{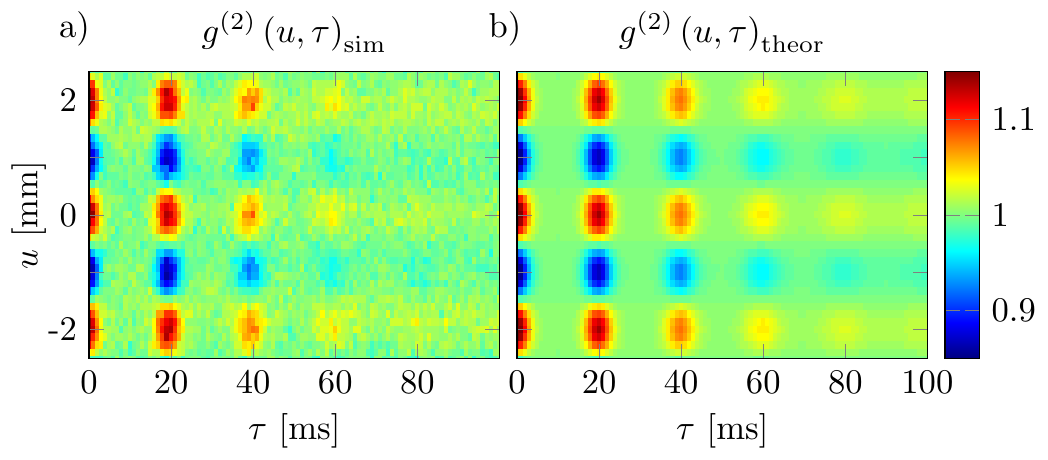} \caption{a) Correlation function of an interference pattern with $K=0.6$ and $\lambda=\unit[2]{mm}$ perturbed by a Gaussian distributed noise according to equation (\ref{eq91}) with $\varphi_0=\unit[2\cdot 10^{-2}]{\pi}$, $\omega_0/2\pi=\unit[50]{Hz}$, $\sigma_\omega/2\pi=\unit[5]{Hz}$ and a frequency resolution of \unit[1]{mHz}. The superperiod of $\tau_s=\unit[20]{ms}$ belongs to the central frequency of $\unit[50]{Hz}$. With the full width at half maximum (FWHM) of the applied noise spectrum being $\text{FWHM}=2\sqrt{2\ln(2)}\sigma_\omega\approx2\pi\times\unit[11.8]{Hz}$, the contrast in the correlation function vanishes on a timescale $\tau=2\pi/\text{FWHM}\approx\unit[85]{ms}$. b) Theoretical correlation function resulting from the fit with equation (\ref{eq92}). Here, the argument of the Bessel function is given by the discrete amplitude spectrum in equation (\ref{eq91}).}
\label{fig12}
\end{figure}

The correlation function of the simulated interference pattern is extracted according to equation (\ref{eq62}) with a spatial and temporal discretization of $\Delta u =$ \unit[67]{\textmu m} and $\Delta\tau =\unit[1]{ms}$, that were chosen to have a good signal in the correlation function. In figure \ref{fig12}(a), the resulting correlation function $g^{(2)}(u,\tau)_{\text{sim}}$ is shown. The contrast $K_{\text{g}^{(2)}}=0.587\pm 0.003$ and spatial periodicity $\lambda_{\text{g}^{(2)}}=\unit[(1.996\pm 0.001)]{mm}$ are determined by fitting equation (\ref{eq36}) to the correlation function at the temporal position $\tau=\unit[0]{ms}$ (see section \ref{sec2.4}). The superperiod of $\tau_s=\unit[20]{ms}$ belongs to the central frequency of $\unit[50]{Hz}$. The contrast of the correlation function decays on timescales of $\tau\propto 2\pi/\sigma_\omega$. In figure \ref{fig12}(a), the contrast is almost vanished for $\tau>\unit[85]{ms}$. This point is shifted to higher correlation times for smaller $\sigma_\omega$ until the single frequency case with $\omega_1/2\pi=\unit[50]{Hz}$ is reached. For larger $\sigma_\omega$, more frequency components with random phases contribute to the perturbation and the resulting time-dependent perturbation $\varphi\left(t_i\right)$ becomes more uncorrelated between two time stamps. Therefore, the corresponding particles are also uncorrelated and the contrast in the correlation function vanishes for shorter correlation times until it is completely lost.

To determine the characteristics of the Gaussian distributed noise ($\varphi_0$, $\omega_0$ and $\sigma_\omega$), equation (\ref{eq34}) and (\ref{eq35}) are used together with the amplitude spectrum of the applied perturbation $\hat{\varphi}\left(\omega_j\right)$ in equation (\ref{eq91}) as argument of the Bessel function
\begin{align}\label{eq92}
g^{(2)}(u,\tau) = 1 + \frac{K_{\text{g}^{(2)}}^2}{2}\cos\left(\frac{2\pi}{\lambda_{\text{g}^{(2)}}} u\right)\cdot \prod_{j=1}^{N_\omega} \Biggl(\sum_{m_j=-\infty}^\infty J_{m_j}\big(\hat{\varphi}\left(\omega_j\right)\big)^2\cos\left(m_j\omega_j\tau\right)\Biggr)~.
\end{align}
The approximate correlation function can be used for the theoretical description of broad-band frequency noise as long as the number of involved frequencies is large enough, so that the constraint in equation (\ref{eq20}) is only fulfilled for $n_j=-m_j$. The contrast $K_{\text{g}^{(2)}}$ and spatial periodicity $\lambda_{\text{g}^{(2)}}$ extracted according to section \ref{sec2.4} are fixed parameters for the fit to the correlation function of the simulation in figure \ref{fig12}(a). The fit parameters are $\varphi_0$, $\omega_0$ and $\sigma_\omega$ in the discrete amplitude spectrum of the Gaussian distributed noise (equation (\ref{eq91})). The resulting theoretical correlation function $g^{(2)}(u,\tau)_{\text{theor}}$ is illustrated in figure \ref{fig12}(b) and shows a good agreement with the correlation function of the simulation. The corresponding amplitude spectrum $\hat{\varphi}\left(\omega_j\right)_{\text{theor}}$ resulting from the fitted theoretical correlation function is plotted in figure \ref{fig11} (red solid line). It is also in good agreement with the amplitude spectrum of the applied perturbation (blue solid line). The determined characteristics of the Gaussian distributed noise in equation (\ref{eq91}) are $\varphi_0=\unit[(1.48\pm 0.02)\cdot 10^{-2}]{\pi}$, $\omega_0/2\pi=\unit[(49.97\pm 0.11)]{Hz}$ and $\sigma_\omega/2\pi=\unit[(5.01\pm 0.11)]{Hz}$, which are in good agreement with the original values.
\begin{figure}
\centering
\includegraphics[width=0.5\textwidth]{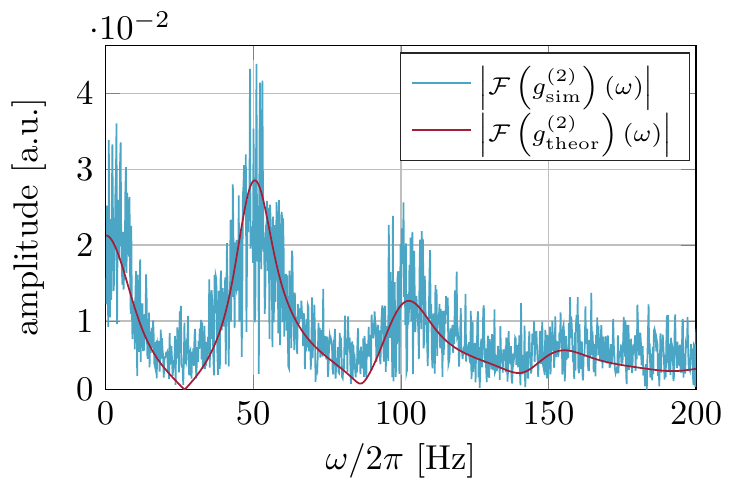} \caption{Amplitude spectra of the simulated (blue solid line) and theoretical correlation function (red solid line) determined from the numerical Fourier transformation calculated at $u=\unit[0]{mm}$.}	
\label{fig13}
\end{figure} 

In figure \ref{fig13}, the amplitude spectra of the simulated and theoretical correlation function, $\left|\mathcal{F}\left(g^{(2)}(0,\tau)_{\text{sim}}\right)(0,\omega)\right|$ (blue solid line) and $\left|\mathcal{F}\left(g^{(2)}(0,\tau)_{\text{theor}}\right)(0,\omega)\right|$ (red solid line), are plotted. Both are calculated with a numerical Fourier transformation at $u=\unit[0]{mm}$ with a frequency resolution of \unit[100]{mHz}. The theoretical amplitude spectrum agrees well with the simulation. The theoretical description of the amplitude spectrum of the correlation function is given by equation (\ref{eq47}) for $u=\unit[0]{mm}$ and positive frequencies using the amplitude spectrum of the perturbation in equation (\ref{eq91}) as argument of the Bessel function
\begin{align}\label{eq93}
\frac{1}{2\pi}\Big|\mathcal{F}\left(g^{(2)}(0,\tau)\right)&(0,\omega)\Big|^2 = \delta(\omega)^2+ \\ \nonumber
&+ \Bigg(\frac{K_{\text{g}^{(2)}}^2}{2}\sum\limits_{\substack{\left\{-m_j,m_j\right\}\in\ker(c)\\ j=1\ldots N_\omega}} \,\left(\prod\limits_{j=1}^{N_\omega} J_{m_j}\big(\hat{\varphi}\left(\omega_j\right)\big)^2\right)\cdot\delta(\omega-\omega_{\{m_j\}})\Bigg)^2 ~. \nonumber
\end{align} 
The broad frequency distribution around \unit[50]{Hz} originates from the fundamental frequencies of the applied perturbation. They are represented in equation (\ref{eq93}) by the first order of the Bessel function $(m_j=1)$. The distributions around \unit[0]{Hz} and \unit[100]{Hz} are generated by the sum and difference frequencies (intermodulation terms) of the perturbation frequencies. Additionally, the distribution around \unit[150]{Hz} originates from the sum of three frequency components of the applied perturbation. 

If the properties of the applied perturbation are not known a priori, the amplitude spectrum can be used to get a reference to the shape and frequency characteristics, because they are included in the spectrum. In figure \ref{fig13}, the central frequency of \unit[50]{Hz} can be identified at the position of the maximum frequency distribution in the amplitude spectrum and used as starting value for the theoretical fit function (equation (\ref{eq92})). If the central frequency of the perturbation would have been \unit[100]{Hz}, for example, the distribution around \unit[50]{Hz} would not be present in the spectrum. The frequency standard deviation of the frequency distribution around \unit[50]{Hz} is broadened because of additional terms in equation (\ref{eq93}), that do not correspond to the fundamental perturbation frequencies $(m_j=1)$. However, it can be used as maximum frequency standard deviation for the theoretical fit function. The applied perturbation spectrum can be identified, if the frequency distributions contained in the resulting amplitude spectrum are separated. They can overlap, if the amplitudes of the perturbation spectrum are large or for very broad spectra.

\subsection{Application limits of the second-order correlation analysis}\label{sec3.5}
The second-order correlation analysis can not be applied under all conditions. Especially, the possibility to determine the perturbation frequency and amplitude is crucial. Therefore, the limits of the applicability of the second-order correlation analysis shall be pointed out.

If the time of flight, that the particles spend in the area of perturbation, is large as compared to the cycle time of the oscillation, it is not possible to resolve the perturbation frequency, because the particles traverse many periods of the oscillation and therefore the perturbation is averaged out. In this case, the correlation analysis can not be applied.

In general, the second-order correlation theory can be used for periodic oscillations, even if the average particle count rate is lower than the perturbation frequency, because of the infinite coherence of such a perturbation. Due to the strong decay of the Bessel function $J_{m_j}(\varphi_j)$, the highest order per perturbation frequency contributing to the correlation function is $m_{j,\text{max}}\approx \varphi_j$. Therefore, the maximum frequency component of all perturbation frequencies included in the correlation function is given by $\text{max}(\varphi_j\omega_j)$, which arises from the argument of the cosine in equation (\ref{eq31}) and (\ref{eq35}). For slow and random perturbations with a high peak phase deviation this product sets a lower limit for the average particle count rate to get a good agreement between experiment and theory.

\section{Conclusion}\label{sec4}
Single-particle interferometry is an outstanding instrument in the field of quantum physics and sensor applications. Due to the high sensitivity of interferometers, they are susceptible to dephasing effects originating from electromagnetic oscillations \cite{Rembold2014,Guenther2015}, mechanical vibrations \cite{Rembold2016} or temperature drifts. Compared to decoherence, dephasing is a collective shift of the particle wave function and the contrast is only reduced in the temporally integrated interference pattern. Therefore, dephasing can in principle be reversed. Using second-order correlation theory, the wave properties can be identified and the perturbation characteristics can be determined. This was demonstrated in former publications for electromagnetic perturbations \cite{Rembold2014,Guenther2015} and mechanical vibrations \cite{Rembold2016}. This paper provides the theoretical fundament for those articles and other future applications in various fields of single-particle interferometry. It gives a detailed description of the analytic solution to the second-order correlation function and its numerical application.

We presented the full analytic derivation of our two-dimensional second-order correlation theory for multifrequency perturbations. The difference between the explicit and approximate solution was discussed and areas of validity were investigated. The amplitude spectra of both solutions, that are used for the identification of the perturbation characteristics \cite{Rembold2016}, have been calculated. We provided the numerical solution of the correlation function and investigated the dependence of the extracted contrast and perturbation amplitude on the discretization step size. The influence of noise on the correlation function and corresponding amplitude spectrum was investigated and an optimum spatial discretization step size was provided to achieve a maximum signal-to-noise ratio. The validity of our calculations could be demonstrated with a single-particle simulation of a perturbed interference pattern evaluated for different spatial discretization step sizes. The possibility to analyze broad-band frequency noise was shown using a simulated interference pattern perturbed by Gaussian distributed noise.

Our method is a powerful tool for the proof of single-particle interferences, even if they are vanished in the spatial signal. Especially for mobile interferometers or experiments in perturbing environments, the requirements for vibrational damping and electromagnetic shielding can be reduced. Furthermore, it is suitable to analyze the characteristics of multifrequency perturbations and broad-band noise. Therefore, it has possible sensor applications, which was demonstrated for mechanical vibrations in an electron interferometer \cite{Rembold2016}. It can be used in principle in every interferometer generating a spatial interference pattern on a detector with high spatial and temporal single-particle resolution. This makes our method applicable in a wide range of experiments.

\section*{Acknowledgments}
We gratefully acknowledge support from the DFG through the SFB TRR21 and the Emmy Noether program STI 615/1-1. A R acknowledges support from the Evangelisches Studienwerk eV Villigst. The authors thank N Kerker and A Pooch for helpful discussions.

\end{document}